%% file: Tx_and_Rx_IQ_imbalance_Compensation_for_OFDM_compliant_JCAS.tex
\documentclass[pdflatex,sn-mathphys-num]{sn-jnl}

\usepackage{graphicx}%
\usepackage{multirow}%
\usepackage{amsmath,amssymb,amsfonts}%
\usepackage{amsthm}%
\usepackage{mathrsfs}%
\usepackage[title]{appendix}%
\usepackage{textcomp}%
\usepackage{manyfoot}%
\usepackage{booktabs}%
\usepackage{algorithmicx}%
\usepackage{listings}%
\usepackage[dvipsnames]{xcolor}
\usepackage{acronym}
\usepackage{algorithm, algpseudocode, float}

\usepackage{tikz, pgfplots}
\usepackage{pgfplotstable}
\usepackage{pbalance}
\usepackage{caption}

\usepackage{siunitx}
\usepackage{makecell} 
\usepackage{subfigure}
\usetikzlibrary{shapes.geometric}
\usepgfplotslibrary{patchplots}
\usetikzlibrary{positioning}
\usetikzlibrary{calc,through,backgrounds}
\usetikzlibrary{shapes.misc}
\usetikzlibrary{spy}
\usetikzlibrary{arrows.meta}

\usepgfplotslibrary{groupplots}

\theoremstyle{thmstyleone}%
\theoremstyle{thmstyletwo}%

\theoremstyle{thmstylethree}%

\begin{document}

\title{Tx and Rx IQ Imbalance Compensation for JCAS in 5G NR}


\author*[1,2]{\fnm{Andreas} \sur{Meingassner}}\email{andreas.meingassner@jku.at}

\author[1]{\fnm{Oliver} \sur{Lang}}\email{oliver.lang@jku.at}

\author[1,2]{\fnm{Moritz} \sur{Tockner}}\email{moritz.tockner@jku.at}

\author[1]{\fnm{Bernhard} \sur{Plaimer}}\email{bernard.plaimer@jku.at}

\author[1]{\fnm{Matthias} \sur{Wagner}}\email{matthias.wagner@jku.at}

\author[2,3]{\fnm{Günther} \sur{Lindorfer}}\email{guenther.lindorfer@jku.at}

\author[2,3]{\fnm{Michael} \sur{Hofstadler}}\email{michael.hofstadler@jku.at}

\author[1,2]{\fnm{Mario} \sur{Huemer}}\email{mario.huemer@jku.at}

\affil*[1]{\orgdiv{Institute of Signal Processing}, \orgname{Johannes Kepler University}, \orgaddress{\street{Altenbergerstraße}, \city{Linz}, \postcode{4040}, \country{Austria}}}

\affil[2]{\orgdiv{Christian Doppler Laboratory for Digitally Assisted RF Transceivers for Future Mobile Communications}, \orgname{Johannes Kepler University}, \orgaddress{\street{Altenbergerstraße}, \city{Linz}, \postcode{4040}, \country{Austria}}}

\affil[3]{\orgdiv{Institute for Communications Engineering and RF-Systems}, \orgname{Johannes Kepler University}, \orgaddress{\street{Altenbergerstraße}, \city{Linz}, \postcode{4040}, \country{Austria}}}


\abstract{Beside traditional communications, \ac{JCAS} is gaining increasing relevance as a key enabler for next-generation wireless systems. The ability to accurately transmit and receive data is the basis for high-speed communications and precise sensing, where a fundamental requirement is an accurate \ac{I} and \ac{Q} modulation. For sensing, imperfections in IQ modulation lead to two critical issues in the \ac{RDM} in form of an increased noise floor and the presence of ghost objects, degrading the accuracy and reliability of the information in the \ac{RDM}.

This paper presents a low-complex estimation and compensation method to mitigate the IQ imbalance effects. This is achieved by utilizing, amongst others, the leakage signal, which is the direct signal from the transmitter to the receiver path, and is typically the strongest signal component in the \ac{RDM}. The parameters of the IQ imbalance suppression structure are estimated based on a mixed complex-/real-valued bilinear filter approach, that considers IQ imbalance in the transmitter and the receiver of the \ac{JCAS}-capable \ac{UE}. The \ac{UE} uses a 5G \ac{NR}-compliant \ac{OFDM} waveform with the system configuration assumed to be predefined from the communication side. To assess the effectiveness of the proposed approach, simulations are conducted, illustrating the performance in the suppression of IQ imbalance introduced distortions in the \ac{RDM}.}


\maketitle

\section{Introduction\label{sec:introduction}}
The global demand for connectivity continuously rises, driven by streaming services, smart devices, as well as the internet of things. Higher data rates and more reliable communication networks are becoming increasingly important. To address these challenges, 5G \ac{NR} was introduced and standardized by the \ac{3GPP} \cite{3gppConfig.38.101-2}. It is designed to provide significantly enhanced performance compared to its predecessor 4G by offering higher data rates, and/or reduced latency within the network. Sensing applications are gaining more relevance due to increased frequency bandwidth available for the \acp{FR} \ac{FR}2-1 and \ac{FR}2-2, ranging from \SI{24.25}{GHz} to \SI{52.6}{GHz} and from \SI{52.6}{GHz} to \SI{71}{GHz}, respectively. The 5G NR standard establishes the bases for \ac{JCAS}, which combines traditional communications with sensing applications \cite{hakobyan2017novel}. \ac{JCAS} can be performed from a \ac{BS} point of view \cite{9363029}, or from the \ac{UE} perspective as investigated in \cite{hofstadler, hakobyan2017novel}. Therein, the transmission node serves not only as communication device, but as a sophisticated sensor node.

A very important point to guarantee accurate transmission and reception of data heavily relies on a precise up- and down-conversion process performed by the transmitter and the receiver, respectively. Accurate IQ modulation and demodulation is crucial as any imbalance between the \ac{I}- and the \ac{Q}-components leads to signal distortions. This so-called IQ imbalance drastically degrades the quality of transmission and reception. The effects of IQ imbalance in 5G NR systems have been extensively investigated in literature, however, mostly from a communications perspective, where IQ imbalance was shown to degrade the \ac{BER}. 

Generally, IQ imbalance estimation methods can be categorized into two different groups. The first group, blind IQ imbalance estimation \cite{6365234,4357467, elsamadouny2014blind, 8755502, 9411715, 10716012} assumes that the data is unknown, but statistical properties of the data are known, serving as the basis for the estimation process. The advantage of these approaches is that they do not require any additional overhead in the data, making it highly efficient for scenarios where bandwidth and resources are limited. However, the accuracy of blind IQ imbalance estimation schemes depends on the quality and quantity of the \ac{Rx} data, which might be a drawback. The second category is pilot-aided or preamble-based IQ imbalance estimation, where a predefined signal known to the receiver is transmitted. It enables high accuracy at the cost of reducing the effective data rate. For the special task of \ac{JCAS}, \ac{Tx} data is transmitted and received at the same device (e.g., \ac{BS} or \ac{UE}), and therefore, all \ac{Tx} data represents the predefined signal known to the receiver. However, the data is random and not specifically designed for IQ imbalance estimation tasks.

In literature, different methods are proposed to mitigate IQ imbalance effects for \ac{JCAS} applications. In \cite{bourdoux2018iq}, an approach is presented to adjust the transmitted waveform in a way that the \ac{OFDM}-based sensing application becomes robust to IQ imbalance. This is done by defining design criterions for the \ac{OFDM} waveform, where the maximum unambiguous range for sensing applications is reduced by a factor of 2, which is, depending on the specific sensing application and data rate, not generally feasible. Based on that, \cite{LangWaveformsJCAS} proposes a novel \ac{OFDM}-based \ac{JCAS} system without this drawback and considering both, \ac{Tx} and \ac{Rx} IQ imbalance. This is achieved by shifting the ghost objects (replicas of objects in the \ac{RDM} caused by IQ imbalance) in the \ac{RDM} to different velocities, where they can be identified by a tracking algorithm. In \cite{schweizer2020iq}, a method is proposed to tackle the issue based on the circularity of the \ac{Rx} signal. This assumption is unfortunately not suitable in presence of strong stationary objects, which motivates a modified IQ imbalance estimation process that eliminates strong stationary objects in the \ac{RDM} as a preprocessing step, and that is also proposed in \cite{schweizer2020iq}. In \cite{ali2016ofdm}, the design of a receiver architecture is studied, which estimates the \ac{CIR} as well as the \ac{Tx} and \ac{Rx} IQ imbalance with a two-step iterative estimation approach. In contrast to other works, the method only requires a short training sequence to estimate the \ac{CIR} as well as the IQ imbalance parameters. Additionally, it is claimed that this iterative method has low computational complexity. In \cite{4024126}, an estimation approach is presented to estimate the IQ imbalance in both, the \ac{Tx} and \ac{Rx} path. It is assumed that the \ac{Tx} IQ imbalance is relatively small compared to the \ac{Rx} IQ imbalance. However, this assumption is in many cases not met in \acp{UE} and restricts the applicability of the method.

Leakage due to direct-coupling from the \ac{Tx} to the \ac{Rx} path is a well known challenge for sensing applications \cite{8805161}. For \ac{FMCW} radar systems this leakage suppression may be performed by down-converting with the \ac{FMCW} transmit signal in the receiver path followed by high-pass filtering \cite{7180336}. Such approaches are not possible for \ac{OFDM} and actual cancellation processes are required to mitigate this effect \cite{8805161}. However, this gives the possibility to exploit the fact that leakage, which is a very dominant signal in the receiver of the \ac{UE}, is present and includes both \ac{Tx} and \ac{Rx} IQ imbalance.

In this work, a method is proposed that utilizes leakage and other dominant signal components at the receiver as input for a mixed complex-/real-valued bilinear estimator proposed in \cite{MSc_Thesis_BP}. The estimated parameters contain information about the IQ imbalance in the \ac{Tx} and \ac{Rx} paths and are subsequently applied in a compensation process, where the true transmission and reception signals are reconstructed to model the effects of IQ imbalance. It is followed by standard \ac{OFDM} radar processing enabling accurate channel estimation. To validate the proposed method, the \acp{RDM} of both the IQ imbalance-affected and the compensated channels are analyzed and the suppression of the increased noise floor and the power reduction of ghost objects are evaluated. Furthermore, the reconstruction of the object's complex amplitudes in the \ac{RDM} is evaluated at the positions of the actual objects to ensure the method's applicability for advanced signal processing tasks, such as, e.g., \ac{DOA} estimation. For this work, a \ac{UE} perspective is assumed, which does not allow neglecting either the \ac{Tx} or \ac{Rx} IQ imbalance as, e.g, done in \cite{4024126}. We assume a 5G \ac{NR}-compliant \ac{OFDM}-system, where the waveform generation is based on a fully allocated \ac{PUSCH} resource grid filled with randomly generated data symbols. In this study, the waveform is considered to be determined by the communication process, making the waveform adjustments proposed in \cite{bourdoux2018iq, LangWaveformsJCAS, schweizer2020iq} unsuitable. The iterative algorithm proposed in \cite{ali2016ofdm} estimating the \ac{CIR} as well as the \ac{Tx} and \ac{Rx} IQ imbalance fulfills all assumptions and is used as a comparison algorithm.

The rest of this paper is organized as follows: In Section~\ref{sec:systemModel}, \ac{OFDM}-based sensing is briefly introduced. The IQ imbalance model, the mathematical derivation of an IQ imbalance afflicted channel, and the corresponding distortions in the \ac{RDM} are discussed in Section~\ref{sec:IQImbalance}. The bilinear filter model, the estimation methods and their behavior are explained in Section~\ref{sec:bilinearFilter}. The compensation strategy is presented in Section~\ref{sec:Compensation}. The parametrization of the 5G \ac{NR}-compliant \ac{OFDM} system, the scenario settings, as well as the simulation results are discussed in Section~\ref{sec:Simulations}. Finally, Section~\ref{sec:Conclusion} concludes this work.
\newline\textit{Notation:}

Vectors are given by lower-case bold face variables $(\mathbf{x}, \mathbf{y}, ...)$, and matrices in the upper-case version $(\mathbf{X}, \mathbf{Y}, ...)$. The transpose, complex-conjugate, and Hermitian operations are denoted by $(\cdot)^{\text{T}}, (\cdot)^*$, and $(\cdot)^{\text{H}}$, respectively. A matrix $\mathbf{X}$, that is complex conjugated and mirrored along the columns is represented by $\underline{\mathbf{X}}^*$ \cite{tubbax2003compensation}, and the element of the $i$th row and the $j$th column of the matrix $\mathbf{X}$ is given by $[\mathbf{X}]_{i,j}$. The real part of a variable $x$ is denoted by $\Re(x)$, and the imaginary part by $\Im(x)$. A real-valued matrix of size $n \times m$ is denoted by $\mathbb{R}^{n \times m}$, and a complex-valued matrix by $\mathbb{C}^{n \times m}$. The element-wise (Hadamard) division is represented by $\oslash$, the element-wise (Hadamard) multiplication by $\circ$, and the Kronecker product by $\otimes$. The element-wise division of a complex conjugated and mirrored vector $\underline{\mathbf{x}}^*$ by the vector itself is denoted with $\mathbf{x}' = \underline{\mathbf{x}}^* \oslash \mathbf{x}$, and the imaginary unit is represented by $j$.

\section{OFDM-Based Sensing\label{sec:systemModel}} 
In \ac{OFDM}-based transmission systems, the complex-valued \ac{Tx} data for all $N_{\text{sc}}$ subcarriers and all $N_{\text{sym}}$ \ac{OFDM} symbols of one subframe is given by the resource grid $\mathbf{X}_{\text{Tx}} \in \mathbb{C}^{N_{\text{sc}} \times N_{\text{sym}}}$. Each \ac{OFDM} symbol in frequency-domain is transformed to the corresponding time-domain signal by applying an \ac{IFFT}, followed by appending a \ac{CP} of duration $T_{\text{CP}}$ to each \ac{OFDM} symbol [4]. The analog time-domain \ac{OFDM} symbol $x(t)$ is obtained by digital to analog conversion and can be mathematically expressed as
\begin{align}
    x(t) = \frac{1}{\sqrt{N_{\text{sc}}}}\sum_{l=0}^{N_{\text{sym}}-1} \sum_{k=0}^{N_{\text{sc}}-1} [\mathbf{X}_{\text{Tx}}]_{k,l}~\text{e}^{j2\pi f_k t_{\text{f}}} ~\text{rect} \left( \frac{t_{\text{f}}+T_{\text{CP}}}{T + T_{\text{CP}}} \right), \label{eq:transmitSignal}
\end{align}
where $f_k = k \Delta f$ represents the frequency of the $k$th \ac{OFDM} subcarrier with a subcarrier spacing of $\Delta f = 1/T$ and with the \ac{OFDM} symbol duration $T$. The (relative) fast-time for the $l$th \ac{OFDM} symbol is $t_{\text{f}} = t - l (T + T_{\text{CP}})$. The rectangular function $\text{rect}(\cdot)$ is defined as 1 when the argument is within [0, 1], and 0 otherwise.
The analog signal comprising multiple \ac{OFDM} symbols is up-converted to the carrier frequency $f_{\text{c}}$, and amplified before transmission. Reflections from objects in the surrounding are received by the \ac{Rx} antenna after the so-called round-trip delays. The received, down-converted complex-valued time-domain signal $y(t)$ without additional noise can thus be modeled as
\begin{align}
    y(t) =& \sum_{i=0}^{N_{\text{path}}-1} \frac{\alpha_i}{\sqrt{N_{\text{sc}}}}\sum_{l=0}^{N_{\text{sym}}-1} \sum_{k=0}^{N_{\text{sc}}-1} [\mathbf{X}_{\text{Tx}}]_{k,l}~\text{e}^{j2\pi f_k (t_{\text{f}}-\tau_i)}~\text{e}^{j2\pi f_{\text{D}_i} t}~\text{rect} \left( \frac{t_{\text{f}}+T_{\text{CP}} - \tau_i}{T + T_{\text{CP}}} \right), \label{eq:receiveSignal}
\end{align}
where the introduced constant amplitude and phase change depending on the $i$th object is represented by $\alpha_i \in \mathbb{C}$, and $\tau_i \in \mathbb{R}$ is the corresponding round-trip delay for $i = 0, 1, \dots, N_{\text{path}}-1$. The Doppler shift regarding the $i$th object is given by $f_{\text{D}_i} = -2 v_i f_{\text{c}}/c_0$, where $v_i \in \mathbb{R}$ represents the relative velocity, and $c_0$ is the speed of light. $N_{\text{path}}$ is the total number of reflections, where each reflection is assumed to correspond to a distinct object in the environment. The received resource grid $\mathbf{Y}_{\text{Rx}} \in \mathbb{C}^{N_{\text{sc}} \times N_{\text{sym}}}$ can be derived after sampling, removing the \ac{CP}, and a subsequent \ac{FFT} for each \ac{OFDM} symbol.

Different approaches exist to derive the channel information as discussed in \cite{10128671}. In this work, the element-wise division also known as zero-forcing filter is used for the \ac{OFDM}-based sensing application, and the frequency-domain channel matrix $\mathbf{H}~\in~\mathbb{C}^{N_{\text{sc}} \times N_{\text{sym}}}$ is computed by
\begin{align}
    \mathbf{H} = \mathbf{Y}_{\text{Rx}}\oslash \mathbf{X}_{\text{Tx}}.
    \label{eq:elementwiseDivision} 
\end{align}
This channel matrix contains a superposition of 2D complex oscillations, one for each reflection in the surrounding. The frequencies of the oscillations are determined by the distance and the relative velocity, which will become important later in this work. The channel matrix $\mathbf{H}$ can be transformed into a \ac{RDM} by applying \acp{IFFT} on all columns and \acp{FFT} on all rows. The \ac{RDM} is a common visualization for the environment of a sensing system, as it illustrates the distance to a reflecting object, as well as its relative velocity.

\section{IQ Imbalance\label{sec:IQImbalance}}

\begin{figure*}
        \centering 
        \input{figures/IQimbalanceBlockDiagram.tex}
        \caption{Block diagram of the IQ imbalance afflicted \acs{JCAS} system in time-domain. First, the \acs{Tx} IQ imbalance is illustrated, followed by a radar channel given by $h(t)$. \acs{AWGN} is added and the second, \acs{Rx} IQ imbalance is modeled subsequently. \label{fig:blockDiagramIQsystem}} 
\end{figure*}
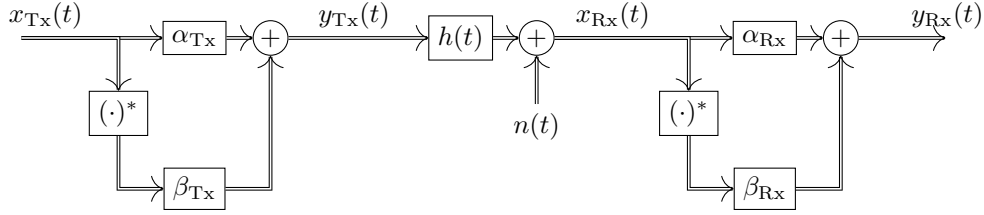

In this section, the employed IQ imbalance model is briefly introduced. The ideal transmit signal $x(t)$, and the imbalance afflicted transmit signal $x_{\text{IQ}}(t)$ are connected via
\begin{align}
    x_{\text{IQ}}(t) &= \alpha x(t) + \beta x^*(t) \label{eq:IQimbalance},
\end{align}
where $\alpha$ and $\beta$ are the \ac{FID} IQ imbalance parameters \cite{tubbax2004compensation, 1204086} calculated by
\begin{align}
    \alpha &= \cos (\Delta \phi/2) + j \epsilon \sin (\Delta \phi/2)\\
    \beta &= \epsilon \cos (\Delta \phi/2) - j \sin (\Delta \phi/2),
\end{align}
with $\epsilon$ and $\Delta \phi$ as the amplitude error and the phase error, respectively. The whole process can also be described in frequency-domain. There, an imbalance afflicted \ac{OFDM} symbol is given by
\begin{align}
    \mathbf{x}_{\text{IQ}}= \alpha \mathbf{x} + \beta \underline{\mathbf{x}}^*,
\end{align} 
where $\mathbf{x}$ is the frequency-domain vector without IQ imbalance and $\underline{\mathbf{x}}^*$ the complex conjugated, mirrored version of it. For this and the following equations, the frequency dependence of a variable is omitted to improve readability.

\subsection{Tx and Rx IQ Imbalance}
The IQ imbalance effects according to the system illustrated in Fig.~\ref{fig:blockDiagramIQsystem} are mathematically derived in this subsection. For better illustration, the IQ imbalance is modeled for each \ac{OFDM} symbol separately. Hence, the IQ imbalance afflicted \ac{Tx} data vector $\mathbf{y}_{\text{Tx}}$ for one \ac{OFDM} symbol is given in frequency-domain by
\begin{align}
    \mathbf{y}_{\text{Tx}} &= \alpha_{\text{Tx}} \mathbf{x}_{\text{Tx}} + \beta_{\text{Tx}} \mathbf{\underline{x}^*_{\text{Tx}}},
\end{align}
where $\mathbf{x}_{\text{Tx}}$ is an arbitrary \ac{OFDM} symbol of the \ac{Tx} resource grid $\mathbf{X}_{\text{Tx}}$, and the \ac{Tx} path IQ imbalance is characterized by the \ac{FID} imbalance parameters $\alpha_{\text{Tx}}$ and $\beta_{\text{Tx}}$. After applying the frequency-domain channel $\mathbf{h}$ and \ac{AWGN} $\mathbf{n}~\sim~\mathcal{CN}(\mathbf{0}, \sigma^2 \mathbf{I}_{N_{\text{sc}}})$ with the noise variance $\sigma^2$ and the identity matrix $\mathbf{I}_{N_{\text{sc}}}$ of size $N_{\text{sc}} \times N_{\text{sc}}$, the \ac{Rx} data vector without \ac{Rx} IQ imbalance is given by
\begin{align}
    \mathbf{x}_{\text{Rx}} &= (\alpha_{\text{Tx}} \mathbf{x}_{\text{Tx}} + \beta_{\text{Tx}} \mathbf{\underline{x}}_{\text{Tx}}^*) \circ \mathbf{h} + \mathbf{n}.
\end{align}
Including the \ac{Rx} IQ imbalance leads to
\begin{align}
    \mathbf{y}_{\text{Rx}} &= \alpha_{\text{Rx}} \alpha_{\text{Tx}} \mathbf{x}_{\text{Tx}} \circ \mathbf{h} + \alpha_{\text{Rx}} \beta_{\text{Tx}} \mathbf{\underline{x}}_{\text{Tx}}^* \circ \mathbf{h}  + \alpha_{\text{Rx}} \mathbf{n} \nonumber \\
    &\qquad + \beta_{\text{Rx}} \alpha_{\text{Tx}}^* \mathbf{\underline{x}}_{\text{Tx}}^*\circ \mathbf{\underline{h}}^* + \beta_{\text{Rx}} \beta_{\text{Tx}}^* \mathbf{x}_{\text{Tx}} \circ \mathbf{\underline{h}}^* + \beta_{\text{Rx}} \mathbf{\underline{n}}^*\label{eq:TRxChannel},
\end{align}
where the \ac{FID} IQ imbalance parameters for the \ac{Rx} path are given with $\alpha_{\text{Rx}}$ and $\beta_{\text{Rx}}$, respectively. 

To retrieve the IQ imbalance afflicted channel vector $\mathbf{h}_{\text{IQ}}$, the \ac{Rx} \ac{OFDM} symbol is divided element-wise by the \ac{Tx} \ac{OFDM} symbol as
\begin{align}
    \mathbf{h}_{\text{IQ}} &= \mathbf{y}_{\text{Rx}} \oslash \mathbf{x}_{\text{Tx}},\\ 
    &= \alpha_{\text{Rx}} \alpha_{\text{Tx}} \mathbf{h} + \alpha_{\text{Rx}} \beta_{\text{Tx}} \mathbf{x}_{\text{Tx}}' \circ \mathbf{h} + \beta_{\text{Rx}} \alpha_{\text{Tx}}^* \mathbf{x}_{\text{Tx}}'\circ \mathbf{\underline{h}}^* \nonumber \\
    &\qquad + \beta_{\text{Rx}} \beta_{\text{Tx}}^* \mathbf{\underline{h}}^* + (\alpha_{\text{Rx}} \mathbf{n} + \beta_{\text{Rx}} \mathbf{\underline{n}}^*)\oslash \mathbf{x}_{\text{Tx}} \label{eq:h_TRx}.
\end{align}

\subsection{Effects of IQ Imbalance on Sensing Applications\label{sec:EffectsOfIQ}}
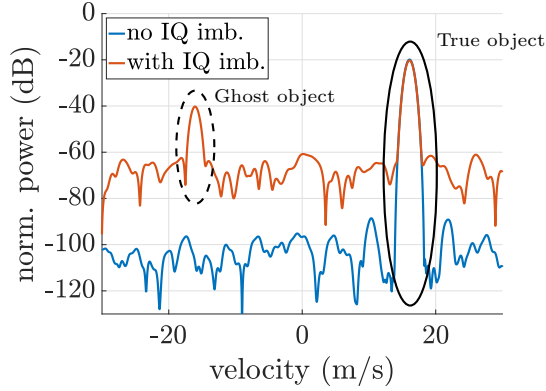
\begin{figure}
        \centering
        \begin{tikzpicture}
                \node (center) [] at (0, 0) {};
                \node at (center) {\includegraphics[width=0.55\linewidth]{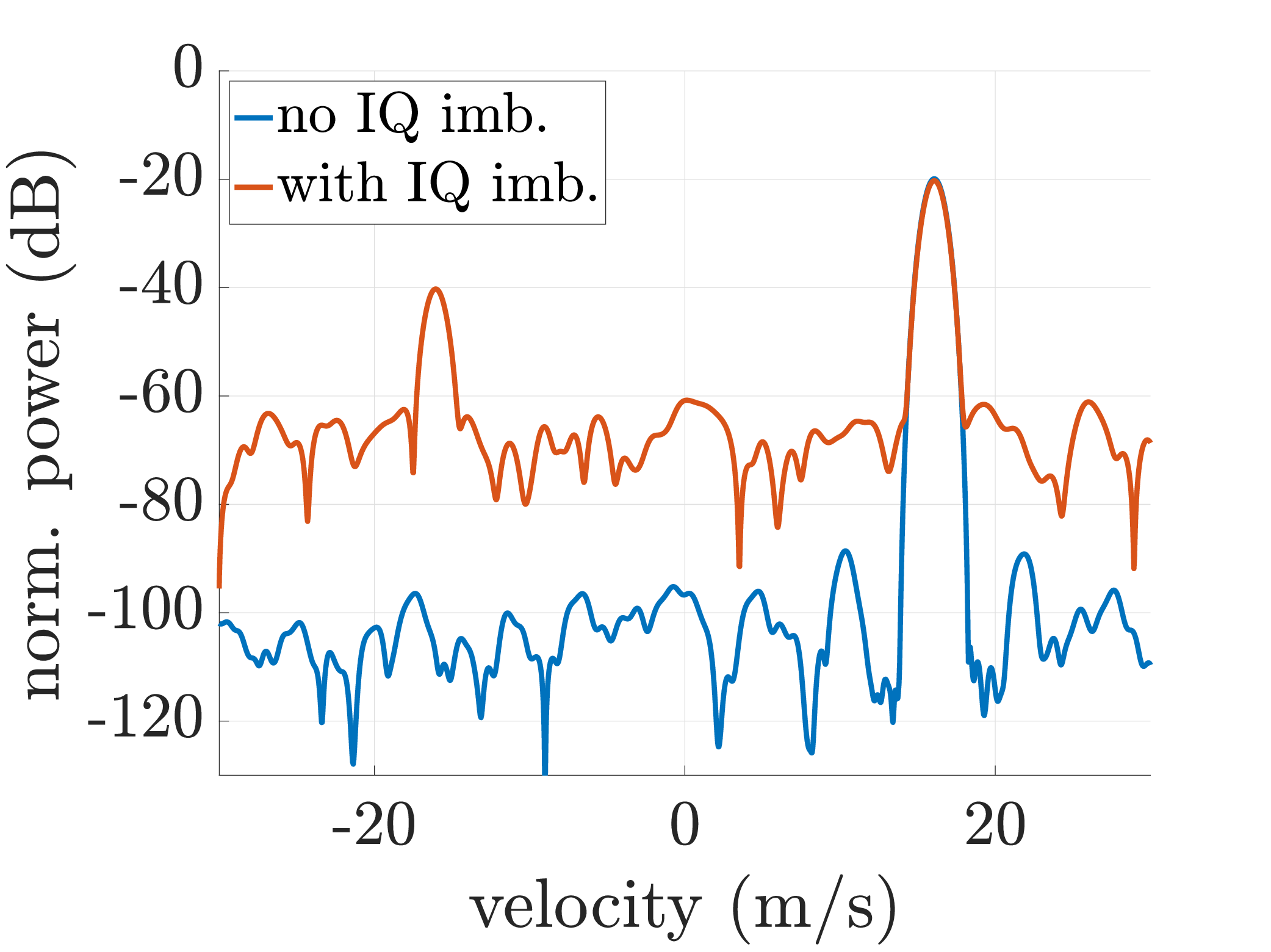}};

                \node (trueObject) [above right=-0.08 and 1.45 of center] {};
                \node[ellipse,draw, minimum width=0.7cm, minimum height=3.5cm, thick] at (trueObject) {};
                \node[above right=1.4cm and 0.1cm of trueObject] {\footnotesize True object};
                
                \node (ghostObject) [above left=0.27 and 0.88 of center] {};
                \node[ellipse,draw, minimum width=0.5cm, minimum height=1.5cm, thick, dashed] at (ghostObject) {};
                \node[above right=0.3cm and 0cm of ghostObject] {\footnotesize Ghost object};

        \end{tikzpicture}
        \caption{Effects of \acs{Tx} and \acs{Rx} IQ imbalance illustrated in the velocity spectrum. The IQ imbalance causes an increased noise floor and ghost objects. \label{fig:IQeffectsSpectrum}}
\end{figure}

When analyzing \eqref{eq:h_TRx}, it can be seen that the first term provides the wanted channel vector $\mathbf{h}$ scaled by a factor $\alpha_{\text{Rx}} \alpha_{\text{Tx}}$ (typically close to 1 under reasonable IQ imbalance assumptions). Additionally, the complex conjugated and mirrored version of the true channel $\mathbf{h}$ scaled by $\beta_{\text{Rx}} \beta_{\text{Tx}}^*$ is present in \eqref{eq:h_TRx}. This term may cause a ghost object at the negative velocity in the \ac{RDM} due to the complex conjugation. Its scaling term is typically close to zero for reasonable IQ imbalance values. Hence, the corresponding peak in the \ac{RDM} can typically be ignored. However, if it is not negligible, this object could easily be identified as ghost object by a tracking algorithm as argued in \cite{LangWaveformsJCAS}, and is therefore not the main focus in this work. Besides the \ac{AWGN} terms, two additional terms depending on $\mathbf{x}_{\text{Tx}}'$ appear, causing an increased noise floor in the \ac{RDM} by \ac{ICI}. Note that the known \ac{Tx} data is the source of this increased noise floor. In Fig.~\ref{fig:IQeffectsSpectrum}, an exemplary velocity spectrum is illustrated based on the aforementioned \ac{OFDM}-based sensing process. The simulation setup imitates a scenario with a dominant leakage at \SI{0}{m} range and \SI{0}{m/s} velocity and a weaker object at \SI{6}{m} and \SI{15}{m/s}. The normalized power levels are \SI{0}{dB} and \SI{-20}{dB}, respectively. The \ac{AWGN} is generated according to a \ac{SNR} of \SI{30}{dB}. These scenario parameters are used for illustration purpose and do not have a real-world reference. They are chosen to particularly highlight the effects of IQ imbalance in the \ac{RDM}. This is also the case for the IQ imbalance parameters. In Fig.~\ref{fig:IQeffectsSpectrum}, the true object, its corresponding ghost object at the negative velocity, and the increased noise floor can be observed, which may cover weaker objects in the \ac{RDM}. Compared to the true object, the ghost object is damped by the so-called \ac{ISR} given by 
\begin{align}
    \text{ISR} &= 10 \log \left(\frac{|\alpha_{\text{Tx}} \alpha_{\text{Rx}}|^2}{|\beta_{\text{Rx}} \beta_{\text{Tx}}^*|^2}\right)
\end{align}
which is of \SI{20}{dB} for this particular parameter setting.

\begin{figure*}
        \centering
        \subfigure[\acs{RDM} according to the channel without IQ imbalance.]{\includegraphics[width=0.49\textwidth]{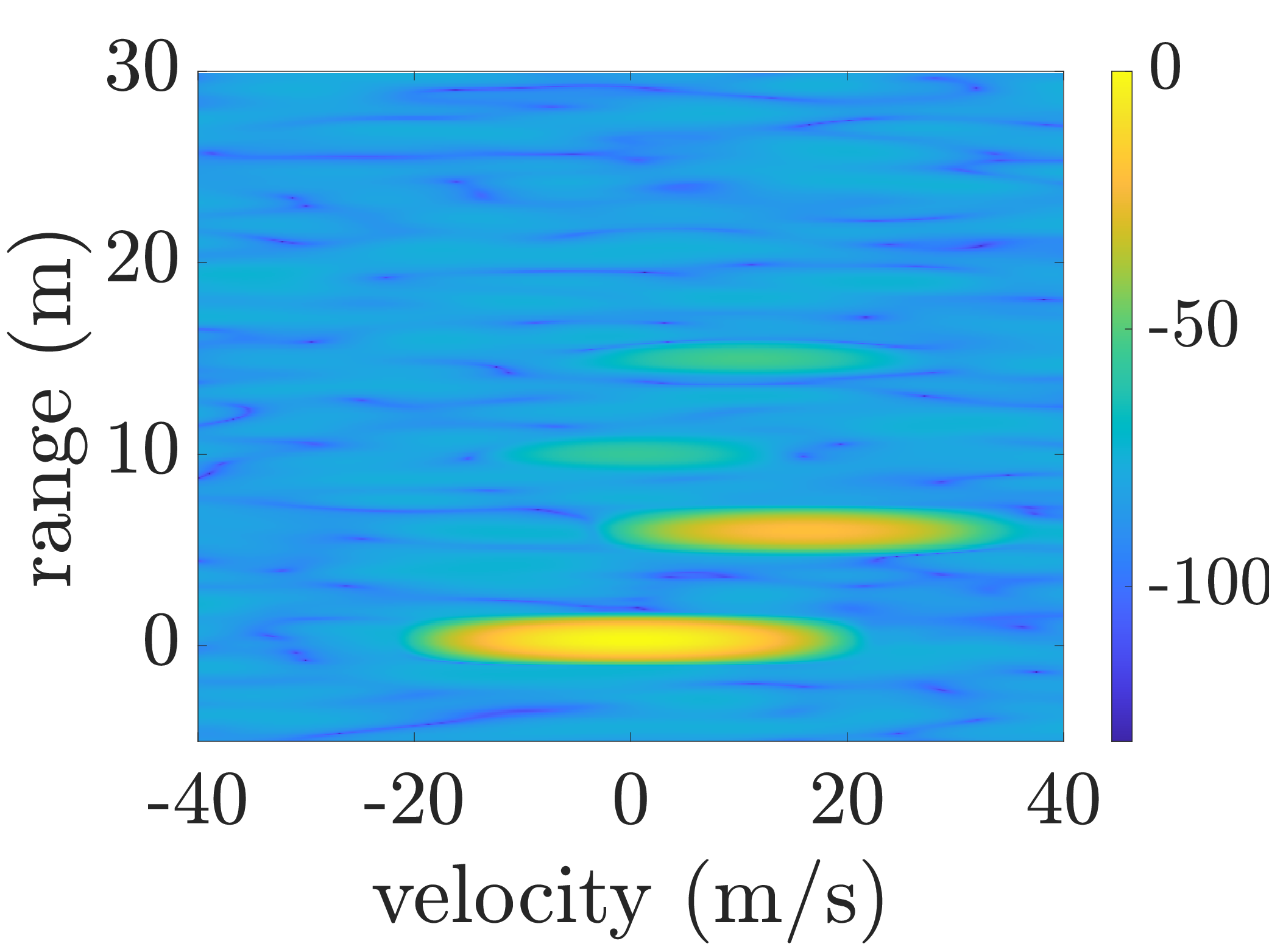}}
        \subfigure[\acs{RDM} according to the IQ imbalance afflicted channel.]{\includegraphics[width=0.49\textwidth]{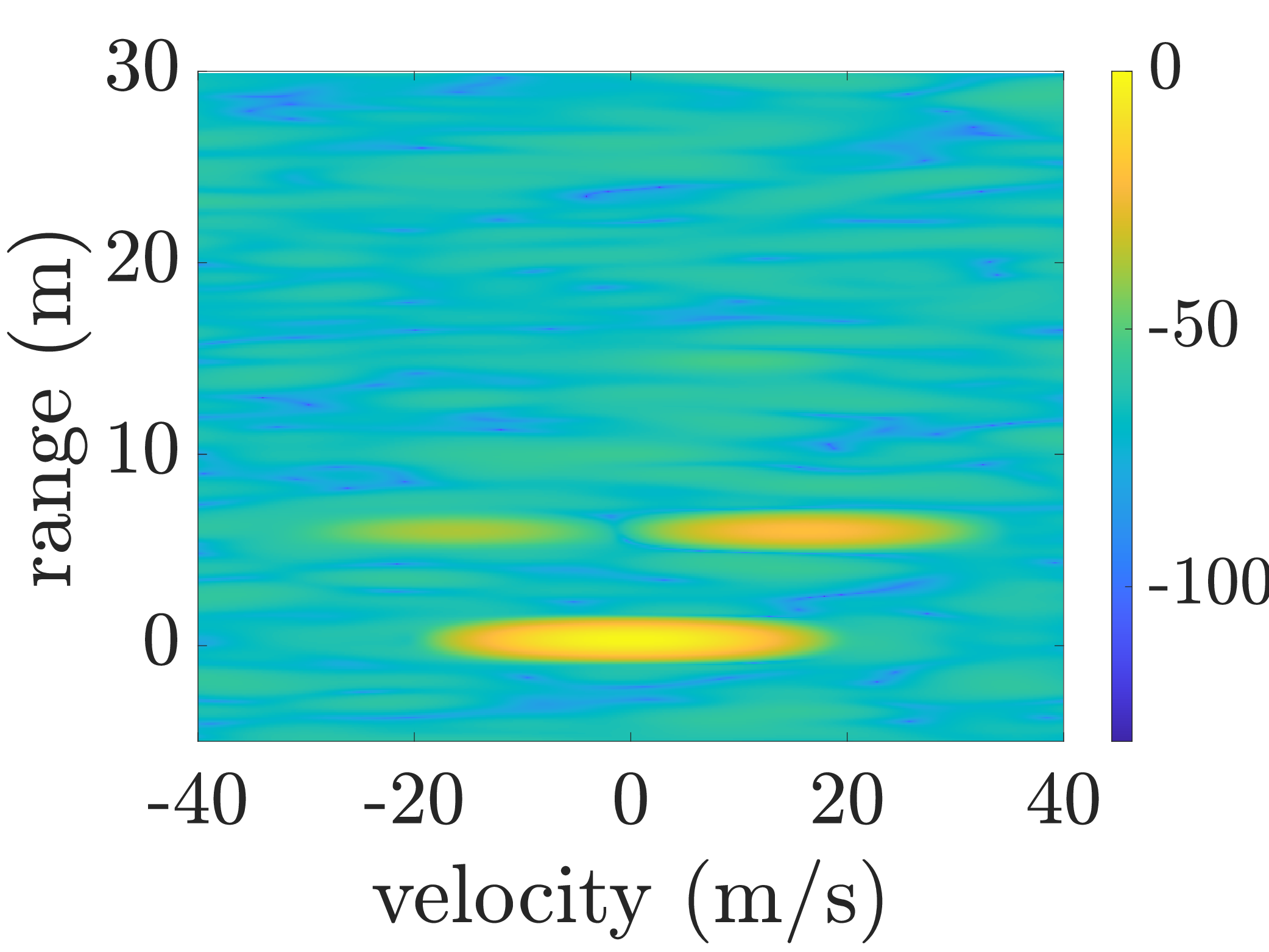}}
        \caption{Effects of \acs{Tx} and \acs{Rx} IQ imbalance illustrated in the \acs{RDM}. The increased noise floor and a ghost object are clearly visible.}
        \label{fig:IQeffectsRDM}
\end{figure*}

In Fig.~\ref{fig:IQeffectsRDM}, an \ac{RDM} according to a) the channel without IQ imbalance, and b) the IQ imbalance afflicted channel is illustrated. For this exemplary scenario (different to the scenario in Fig.~\ref{fig:IQeffectsSpectrum}), a dominant leakage and three additional weaker objects are present. In Fig.~\ref{fig:IQeffectsRDM}b, the increased noise floor as well as the ghost object due to IQ imbalance can be clearly observed.

\section{Parameter Estimation via Bilinear Filters\label{sec:bilinearFilter}}
In \cite{MSc_Thesis_BP}, complex-valued bilinear filters are presented that form the basis for the estimation methods derived in this work. In Section~\ref{sec:EffectsOfIQ}, different terms of \eqref{eq:h_TRx} are discussed. It was observed, that the described \ac{ICI} is the basis for the increased noise floor. The most dominant signal components have the main influence on this effect. Hence, in this work the IQ imbalance parameters $\alpha_{\text{Tx}}, \beta_{\text{Tx}}, \alpha_{\text{Rx}}$, and $\beta_{\text{Rx}}$ are estimated based on a synthetically generated channel matrix corresponding to the most dominant oscillations. Especially when considering a \ac{UE}-centric \ac{JCAS} system, the most dominant reflection is typically the internal leakage, which originates, e.g., from insufficient antenna isolation. In this work, this non-ideal behavior is explicitly exploited to get a first estimate of the channel matrix $\mathbf{H}$ which is denoted by $\tilde{\mathbf{H}}$ and found by the following steps.
\begin{itemize}
    \item Derive the IQ imbalance afflicted channel matrix $\mathbf{H}$ according to \eqref{eq:elementwiseDivision}.
    \item Apply a \ac{CFAR} algorithm to identify dominant peaks that well-exceed the noise floor.
    \item Ignore peaks that may be the ghost object of a stronger peak for the following considerations.
    \item Estimate the distances, relative velocities, and the complex amplitudes of the considered peaks. In this work, the RELAX method \cite{Meingassner,liu1998implementation} is considered for this task due to its accuracy and low computational complexity. 
    \item Use these estimates to synthetically reconstruct the channel matrix $\tilde{\mathbf{H}}$.
\end{itemize}
In the following, the derivation and the behavior of the bilinear filters is presented.

\subsection{Bilinear Model\label{sec:bilinearFilterModel}}
\begin{figure}
    \centering
    \footnotesize
    \begin{tikzpicture}[every node/.style=on grid]
        \node (center) [] at (0, 0) {};

        \node (input) [left=8 of center] {};
        \node (centerTx) [left=2.5 of center] {}; 
        \node (centerRx) [below left=2.4 and 4 of center] {}; 

        \node (tempInput1) [above left=0.6 and 4 of center] {}; 
        \node (tempInput11) [above left=0.6 and 3.7 of center] {}; 
        \node (tempInput2) [above left=0.2 and 7.4 of center] {}; 
        \node (tempInput22) [above left=0.5 and 7.4 of center] {}; 
        \node (tempInput3) [below left=0.2 and 7.4 of center] {}; 
        \node (tempInput33) [below left=0.5 and 7.4 of center] {}; 

        \node (IFFT) [left=3.9 of centerTx, draw] {IFFT};
        \node (DAC) [left=2.6 of centerTx, draw] {DAC};
        \node (AddCP) [left=1.5 of centerTx, draw] {CP};
        \node (betaTx) [below=1.2 of centerTx, draw] {$\beta_{\text{Tx}}$};
        \node (alphaTx) [right=0 of centerTx, draw] {$\alpha_{\text{Tx}}$};
        \node (conjTx) [above left=0.6 and 0.8 of betaTx, draw] {$(\cdot)^*$};
        \node (adderTx) [right=0.9 of centerTx, draw, circle, inner sep=1pt] {$+$};
        \node (channel) [left=-2 of centerTx, draw] {$h(t)$};

        \node (tempChannel1) [below left=1 and -2.8 of centerTx]{};
        \node (tempChannel11) [below left=1.3 and -2.8 of centerTx]{};
        \node (tempChannel2) [below left=1.8 and 3 of center]{};
        \node (tempChannel22) [below left=1.8 and 3.3 of center]{};
        \node (tempChannel3) [above right=0.4 and -2.2 of centerRx]{};
        \node (tempChannel33) [above right=0.1 and -2.2 of centerRx]{};

        \node (adder) [right=-1.5 of centerRx, draw, circle, inner sep=1pt] {$+$};
        \node (noise) [below right=0.7 and -1.5 of centerRx] {}; 
        \node (betaRx) [below=1.2 of centerRx, draw] {$\beta_{\text{Rx}}$};
        \node (alphaRx) [right=0 of centerRx, draw] {$\alpha_{\text{Rx}}$};
        \node (conjRx) [above left=0.6 and 0.8 of betaRx, draw] {$(\cdot)^*$};
        \node (adderRx) [right=0.9 of centerRx, draw, circle, inner sep=1pt] {$+$};
        \node (RemCP) [right=1.5 of adderRx] {};

        \node (RemCP) [right=1.8 of centerRx, draw] {CP};
        \draw[line width=0.05mm] (RemCP.north west) -- (RemCP.south east);
        \draw[line width=0.05mm] (RemCP.north east) -- (RemCP.south west);
        \node (ADC) [right=2.9 of centerRx, draw] {ADC};
        \node (FFT) [right=4.1 of centerRx, draw] {FFT};
        \node (elemDiv) [right=5.1 of centerRx, draw] {$\oslash$};

        \node (bilinearFilter) [below right=2 and 3 of centerRx, draw, minimum height=0.7cm, minimum width=1cm, fill=white] {$\hat{\mathbf{f}}, \hat{\mathbf{g}}$};
        \node (bilinearFilterText) [above=0.8 of bilinearFilter] {Bilinear Filter};
        \node (hEst) [] at (-2.5,-4.6) {};

        \node (adderError) [below right=2 and 5.1 of centerRx, draw, circle, inner sep=1pt] {$+$};

        \node (outputTemp1) [below right=2 and 5.6 of centerRx] {};
        \node (outputTemp11) [below right=2.3 and 5.6 of centerRx] {};
        \node (outputTemp2) [below right=2.8 and 4.5 of centerRx] {};
        \node (outputTemp22) [below right=2.8 and 4.2 of centerRx] {};
        \node (outputTemp3) [below right=2.8 and 2.5 of centerRx] {};
        \node (outputTemp4) [below right=1.25 and 3.45 of centerRx] {};

        \node (output) [below right=2 and 6.1 of centerRx] {};

        \draw [->, double] (input) -- (IFFT) node [at start, above] {$\mathbf{x}_{\text{Tx}}$};
        \draw [->, double] (IFFT) -- (DAC);
        \draw [->, double] (DAC) -- (AddCP);

        \draw [->, double] (AddCP) -- (alphaTx);
        \draw [->, double] (alphaTx) -- (adderTx);
        \draw [->, double] (AddCP) -| (conjTx);
        \draw [->, double] (conjTx) |- (betaTx);
        \draw [->, double] (betaTx)  -| (adderTx);
        \draw [->, double] (adderTx) -- (channel);

        \draw [double] (channel) -| (tempChannel11);
        \draw [double] (tempChannel1) |- (tempChannel22);
        \draw [double] (tempChannel2) -| (tempChannel33);
        \draw [double] (tempChannel3) |- (adder);

        \draw [->, double] (noise) -- (adder) node [at start, below] {$n(t)$};

        \draw [->, double] (adder) -- (alphaRx);
        \draw [->, double] (alphaRx) -- (adderRx);
        \draw [->, double] (adder) -| (conjRx);
        \draw [->, double] (conjRx) |- (betaRx);
        \draw [->, double] (betaRx)  -| (adderRx);
        \draw [->, double] (adderRx) -- (RemCP);
        
        \draw [->, double] (RemCP) -- (ADC);
        \draw [->, double] (ADC) -- (FFT);
        \draw [->, double] (FFT) -- (elemDiv);
        
        \draw [double] (input) -| (tempInput22);
        \draw [double] (tempInput2) |- (tempInput11);
        \draw [->, double] (tempInput1) -| (elemDiv);

        \draw [double] (input) -| (tempInput33);
        \draw [->, double] (tempInput3) |- (-1.5,-4.2);
        \draw [->, double] (hEst) -- (-1.5,-4.6) node [at start, left] {$\hat{\mathbf{h}}$};

        \draw [->, double] (bilinearFilter) -- (adderError) node [at end, below] {-};
        \draw [->, double] (bilinearFilter) -- (adderError) node [midway, above] {$\hat{\mathbf{h}}_{\text{IQ}}$};
        \draw [->, double] (elemDiv) -- (adderError) node [midway, right] {$\mathbf{h}_{\text{IQ}}$};
        \draw [->, double] (adderError) -- (output) node [at end, above] {$\mathbf{e}$};

        \draw [dashed, double] (adderError) -| (outputTemp11);
        \draw [dashed, double] (outputTemp1) |- (outputTemp22);
        \draw [dashed, double] (outputTemp2) -- (outputTemp3);
        \draw [->, dashed, double] (outputTemp3) -- (outputTemp4);

        \node (bilinearFilter) [below right=2 and 3 of centerRx, draw, minimum height=0.7cm, minimum width=1cm, fill=white] {$\hat{\mathbf{f}}, \hat{\mathbf{g}}$};

    \end{tikzpicture}
    \caption{Block diagram of the real system and the bilinear filter to estimate the parameters $\hat{\mathbf{f}}$ and $\hat{\mathbf{g}}$ for subsequent IQ imbalance effect compensation. Note, the sample adaptive filters update the parameters $\hat{\mathbf{f}}$ and $\hat{\mathbf{g}}$ according to the error $\mathbf{e}$, which is illustrated by the dashed feedback line. Additionally, the investigated methods update the estimation parameters based on the $i$th element of estimation error $\mathbf{e}$, which is not illustrated here to keep the block diagram simple.\label{fig:realSystem}} 
\end{figure}

The IQ imbalance afflicted channel $\mathbf{h}_{\text{IQ}}$ from \eqref{eq:h_TRx} can be rewritten to match the bilinear model in \cite{MSc_Thesis_BP} with
\begin{align}
    h_{\text{IQ}, i} &= \mathbf{f}^{\text{H}} \mathbf{X}_i \mathbf{g} + \tilde{n}_i, \label{eq:bilinearForm}
\end{align}
where $\mathbf{f}$ and $\mathbf{g}$ are the parameter vectors, the noise is represented by $\tilde{n}_i$, and $\mathbf{X}_i$ is the known system matrix with matching dimensions. One obtains 
\begin{align}
    &h_{\text{IQ}, i}=
    \underbrace{
        \begin{bmatrix}
            \alpha_{\text{Rx}} & \beta_{\text{Rx}}
        \end{bmatrix}
    }_{\mathbf{f}^{\text{H}}}
    \underbrace{\begin{bmatrix}
            \tilde{h}_i&j\tilde{h}_i&\tilde{h}_i [\mathbf{x}_{\text{Tx}}']_i&j\tilde{h}_i [\mathbf{x}_{\text{Tx}}']_i\\
            [\underline{\tilde{\mathbf{h}}}^*]_i [\mathbf{x}_{\text{Tx}}']_i&-j[\underline{\tilde{\mathbf{h}}}^*]_i [\mathbf{x}_{\text{Tx}}']_i&[\underline{\tilde{\mathbf{h}}}^*]_i&-j[\underline{\tilde{\mathbf{h}}}^*]_i
        \end{bmatrix} 
    }_{\mathbf{X}_i}
    \underbrace{
        \begin{bmatrix}
            \Re(\alpha_{\text{Tx}})\\
            \Im(\alpha_{\text{Tx}})\\
            \Re(\beta_{\text{Tx}})\\
            \Im(\beta_{\text{Tx}})
        \end{bmatrix}
    }_{\mathbf{g}}
    + \tilde{n}_i
    ,\label{eq:hTRxMatrixForm}
\end{align}
where the parameter vector $\mathbf{f} \in \mathbb{C}^2$, and the parameter vector $\mathbf{g} \in \mathbb{R}^{4}$ are corresponding to the \ac{Rx} and \ac{Tx} IQ imbalance parameters, respectively. The noise term $\tilde{n}_i$ is representing $(\alpha_{\text{Rx}} n_i + \beta_{\text{Rx}} [\mathbf{\underline{n}}^*]_i)\oslash \mathbf{x}_{\text{Tx}}$, and is neglected for the subsequent derivation since it does not change the algorithm and only makes the equations longer. However, the term is considered in all simulations later in this work.

For the bilinear model in \eqref{eq:hTRxMatrixForm} with real-valued $\mathbf{g}$ and complex-valued $\mathbf{f}$, several applicable bilinear filters are derived in \cite{MSc_Thesis_BP}, including the bilinear \ac{LMS} filter, the bilinear \ac{NLMS} filter, the bilinear \ac{AWF}, the bilinear \ac{IWF}, as well as the bilinear \ac{RLS} filter. These filters have in common that they iteratively estimate $\mathbf{f}$ for a fixed $\hat{\mathbf{g}}$, followed by an estimation of $\mathbf{g}$ for a fixed $\hat{\mathbf{f}}$. Compared to the sample adaptive bilinear filters \ac{LMS}, \ac{NLMS}, and \ac{RLS}, the Wiener filters perform iteratively. This is exemplarily shown by the update equations for the bilinear \ac{LMS} filter with 
\begin{align}
    \hat{\mathbf{f}}_{i} &= \hat{\mathbf{f}}_{i-1} + \mu_{\text{f}} e_i^* \mathbf{X}_i \hat{\mathbf{g}}_{i-1},
\end{align}
and
\begin{align}
    \hat{\mathbf{g}}_{i} &= \hat{\mathbf{g}}_{i-1} + \mu_{\text{g}} 2 \Re\left\{e_i \mathbf{X}^{\text{H}}_i \hat{\mathbf{f}}_{i-1}\right\},
\end{align}
where $\mu_f$ and $\mu_g$ are the step-sizes, and $e_i = h_{\text{IQ}, i} - \hat{\mathbf{f}}_{i-1}^{\text{H}} \mathbf{X}_i \hat{\mathbf{g}}_{i-1}$ is the error between the $i$th element of the IQ imbalance afflicted channel $\mathbf{h}_{\text{IQ}}$ and the result yielding from the bilinear model corresponding to the $i$th estimate.

\subsection{Computational Complexity of the Bilinear Filter Approaches}
Besides the estimation accuracy, the computational complexity of the different filter approaches is a crucial aspect, especially from the \ac{UE} point of view. Therefore, in this section the number of multiplications, additions and divisions is analyzed. For a multiplication of two complex-valued numbers 4 real-valued multiplications, and 2 real-valued additions are assumed. This aims to provide a suitable comparison to the literature, although there exist other methods, requiring, e.g., 3 real-valued multiplications and 5 real-valued additions \cite{compMul}. For an addition of two complex-valued numbers, 2 real-valued additions are assumed, while for the complex-valued division 6 real-valued multiplications, 3 real-valued additions, and 2 real-valued divisions are considered \cite{cariow2016algorithm}.

\begin{table}[!t]
    \centering
    \caption{Computational complexity of matrix operations with $A \in \mathbb{R}^{L\times M}$, $B \in \mathbb{R}^{M\times N}$, and $C \in \mathbb{R}^{L\times L}$.}
    \begin{tabular}{lccc}
    \hline
        \toprule
        & \textbf{additions} & \textbf{multiplications} & \textbf{divisions} \\
        \midrule
        $A \cdot B \in \mathbb{R}^{L \times N}$ & $L\cdot N\cdot(M-1)$ & $L \cdot N \cdot M$ & 0 \\ 
        \midrule
        $A \otimes B \in \mathbb{R}^{LM\times MN}$ & 0 & $ L \cdot M \cdot M \cdot N$ & 0 \\
        \midrule
        $C^{-1} \in \mathbb{R}^{L \times L}$ & $\frac{2L^3+3^2-5}{6}$ & $\frac{2L^3+3^2-5L}{6}$ & $\frac{L^2-L}{2}$ \\
        \midrule
        $A^*, A^{\text{T}}, A^{\text{H}}$  &  0 & 0 & 0 \\ 
        \bottomrule
    \end{tabular}
    \label{tab:MatrixOperation_computationalComplexity}
\end{table}

For different matrix operations the number of multiplications, additions, and divisions are calculated and presented in Table~\ref{tab:MatrixOperation_computationalComplexity}, where it is assumed that the matrices are real-valued. The matrix inversion illustrated in line 3 is given according to \cite{gowers2010true}. For complex-valued matrix operations the aforementioned assumptions regarding the number of real-valued multiplications, additions and divisions have to be considered. 

\begin{table*}[!t]
    \centering
    \footnotesize
    \caption{Computational complexity per iteration of different mixed complex-/real-valued bilinear filter approaches proposed in \cite{MSc_Thesis_BP} with $\mathbf{f} \in \mathbb{C}^{L}$ and $\mathbf{g} \in \mathbb{R}^{M}$. Additionally, the computational complexity to calculate the statistics $\hat{\mathbf{R}}_{\mathbf{xx}}$ and $\hat{\mathbf{R}}_{\mathbf{Xy}}$  is listed with $N$ as the number of averaged values.}
    \begin{tabular}{lcc}
    \hline
        \toprule
        & \textbf{additions} & \textbf{multiplications}\\
        \midrule
        AWF
            & \makecell{$ \frac{4 L^3 (3 M^2+3 M+2)}{6} + \frac{3 L^2 (8 M^3+4 M+7)}{6}$ \\ $ +\frac{L (24 M^3-12 M^2+12 M-17)}{6}+\frac{2 M^3+3 M^2-17M}{6}$} 
            & \makecell{$ \frac{4 L^3 (3 M^2+3 M+1)}{6}+ \frac{3 L^2 (8 M^3+8 M+10)}{6} $ \\ $ + \frac{L (24 M^3+24 M-2)}{6} +\frac{2 M^3+9 M^2-5M}{6}$} \\ 
        \midrule
        IWF 
            & \makecell{$2 L^3 (M^2+M)+2L^2 \left(2 M^3- M\right) $\\$+L \left(4 M^3-2M^2+4 M+1\right)-M^2$ }
            & \makecell{$2 L^3 (M^2+M)+2L^2 \left(2 M^3+2\right)$\\$+L \left(4 M^3+4 M+2\right)+M^2 $}\\
        \midrule
        LMS 
            & $4ML+6L +6$
            & $4LM +12L+M+18$\\
        \midrule
        NLMS  
            & $6LM +4L+3M+9$ 
            & $6LM +12L+2M+19$\\ 
        \midrule
        RLS  
            & $4L^2+14M^2+6ML+12L+23M+25$ 
            & $10L^2+14M^2+6ML+10L+30M+31$ \\ 
        \midrule
        $\hat{\mathbf{R}}_{\mathbf{xx}} \& \hat{\mathbf{R}}_{\mathbf{Xy}}$ 
            & $2 \left(L^2 M^2 N+L M N+4\right)$
            & $6 L^2 M^2+6 L M+16$\\ 
        \bottomrule
    \end{tabular}
    \label{tab:BilinearFilters_computationalComplexity}
\end{table*}
With that, the computational complexities per iteration of the mixed complex-/real-valued bilinear filter approaches are listed in Table~\ref{tab:BilinearFilters_computationalComplexity} with $\mathbf{f} \in \mathbb{C}^{L}$ and $\mathbf{g} \in \mathbb{R}^{M}$. In addition to Table~\ref{tab:BilinearFilters_computationalComplexity}, the number of real-valued divisions of the \ac{AWF} are given with $L^2+L + (M^2 + M)/2$, for the \ac{NLMS} filter with 2, for the  filter with 9, and for the statistics $\hat{\mathbf{R}}_{\mathbf{xx}}$ and $\hat{\mathbf{R}}_{\mathbf{Xy}}$ (required for the Wiener filters) with 1. The other approaches do not require divisions at all. For the model in \eqref{eq:hTRxMatrixForm} one can obtain $L=2$ and $M=4$. With that, the numbers of required real-valued multiplications, additions and divisions are listed in Table~\ref{tab:BilinearFilters_AppliedComputationalComplexity}.

\begin{table}[!t]
    \centering
    \caption{Number of required real-valued  multiplications, additions, and divisions according to Table~\ref{tab:BilinearFilters_computationalComplexity} adopted for the model in \eqref{eq:hTRxMatrixForm} with $\mathbf{f} \in \mathbb{C}^\text{2}$ and with $\mathbf{g} \in \mathbb{R}^\text{4}$.}
    \begin{tabular}{lccc}
        \hline
            \toprule
            & \textbf{additions} & \textbf{multiplications} & \textbf{divisions} \\
            \midrule
            AWF & 1877 & 2024 & 16 \\ 
            \midrule
            IWF & 1778 & 1916 & 0 \\
            \midrule
            LMS & 50 & 78 & 0 \\
            \midrule
            NLMS & 77 & 99 & 2 \\ 
            \midrule
            RLS & 429 & 483 & 9 \\ 
            \midrule
            $\hat{\mathbf{R}}_{\mathbf{xx}} \& \hat{\mathbf{R}}_{\mathbf{Xy}}$ & 8 + 144 N & 448 & 1 \\ 
            \bottomrule
        \end{tabular}
    \label{tab:BilinearFilters_AppliedComputationalComplexity}
\end{table}

It can be seen that the \ac{AWF} and the \ac{IWF} are very computational complex, however, these approaches do not need a huge number of iterations to estimate the parameters. In contrast to the sample adaptive bilinear filters \ac{LMS}, \ac{NLMS}, and \ac{RLS}, the Wiener filters require estimates of the statistics $\mathbf{R}_{\mathbf{xx}} = \text{E}[\mathbf{x}_k\mathbf{x}_k^{\text{H}}] \in \mathbb{C}^{ML \times ML}$, and $\mathbf{R}_{\mathbf{Xy}} = \text{E}[\mathbf{X}_k \mathbf{y}_k^*] \in \mathbb{C}^{L \times M}$. In this work these estimates are derived by
\begin{align}
    \hat{\mathbf{R}}_{\mathbf{xx}} &= \frac{1}{N} \sum_{k=1}^{N} \mathbf{x}_k \mathbf{x}_k^{\text{H}}\ \qquad \text{with}~\mathbf{x}_k = \text{vec}(\mathbf{X}_k)\\
    \hat{\mathbf{R}}_{\mathbf{Xy}} &= \frac{1}{N} \sum_{k=1}^{N} \mathbf{X}_k y_k^*,
\end{align}
where $\text{vec}(\cdot)$ represents the vectorization operation, $N$ is the number of averaged values, and $y_k$ is the output of the bilinear filter, which equals $ h_{\text{IQ}, k}$. The corresponding computational complexity is given in the last line of Table~\ref{tab:BilinearFilters_computationalComplexity} for the general case, and in the last line of Table~\ref{tab:BilinearFilters_AppliedComputationalComplexity} when applied to the model in \eqref{eq:hTRxMatrixForm}.

The same analysis is done for the iterative algorithm proposed in \cite{ali2016ofdm}. Compared to the bilinear filter approaches, this algorithm processes the time-domain signal as estimation input. Hence, one \ac{OFDM} symbol without \ac{CP} of length $N_{\text{FFT}} = 4096$ (given by the 5G NR standard) is used for the estimation process, resulting in
\begin{itemize}
    \item Number of real-valued additions  \\
    $\qquad \frac{128}{3}N_{\text{FFT}}^3 + 42 N_{\text{FFT}}^2 + \frac{169}{3} N_{\text{FFT}} + 43 \thickapprox 2.93 \cdot 10^{12}$,
    \item Number of real-valued multiplications \\
    $\qquad \frac{128}{3}N_{\text{FFT}}^3 + 54 N_{\text{FFT}}^2 + \frac{247}{3} N_{\text{FFT}} + 85 \thickapprox 2.93 \cdot 10^{12}$,
    \item Number of real-valued divisions \\
    $\qquad 2 N_{\text{FFT}}^2 + N_{\text{FFT}} + 21 \thickapprox 3.36 \cdot 10^{7}$.
\end{itemize}
It can be seen, that the computational complexity is drastically increased compared to the bilinear approaches, however, one \ac{OFDM} symbol is used for one iteration. The proposed method is applied as given in \cite{ali2016ofdm} without advanced computational complexity optimizations.

\subsection{Ambiguity of the Estimation Process\label{sec:EstimationBasedOnTrueChannel}}
Using the bilinear model of \eqref{eq:bilinearForm}, it can be seen that for a given $\mathbf{X}_i$, the estimates $\hat{\mathbf{f}}$ and $\hat{\mathbf{g}}$ are not unique. This can be illustrated by extending the equation as 
\begin{align}
    h_{\text{IQ}, i} &= \lambda \hat{\mathbf{f}}^{\text{H}} \mathbf{X}_i \hat{\mathbf{g}} \frac{1}{\lambda^*}, \label{eq:trivialBilinearSolution}
\end{align}
where $\lambda \in \mathbb{C}$ is an arbitrary complex factor fulfilling $\lambda \neq 0$. Hence, the same output of the bilinear model can be described by a scaled $\hat{\mathbf{f}}$ and inversely scaled $\hat{\mathbf{g}}$. Interestingly, for the derived bilinear model in \eqref{eq:hTRxMatrixForm} it turns out that not one but two closely related scaling factors can be identified, which is shown in the following. For this derivation 
\eqref{eq:hTRxMatrixForm} is given with general matrix and vector elements $x_i, f_i$ and $g_i$ to limit the length of the equations 
\begin{align}
    h_{\text{IQ}, i}&=
    \begin{bmatrix}
        f_1^* & f_2^*
    \end{bmatrix}
    \begin{bmatrix}
        x_1 &  jx_1 & x_2 &  jx_2 \\
        x_3 & -jx_3 & x_4 & -jx_4
    \end{bmatrix} 
    \begin{bmatrix}
        \Re(g_1)\\
        \Im(g_1)\\
        \Re(g_2)\\
        \Im(g_2)
    \end{bmatrix}\label{eq:generalMatrixForm}.
\end{align}
By scaling $f_1^*$ with $\lambda_1$ and $f_2^*$ with $\lambda_2$, \eqref{eq:generalMatrixForm} can be written as 
\begin{align}
    h_{\text{IQ}, i}&=
    \begin{bmatrix}
        \lambda_1 f_1^* & \lambda_2 f_2^*
    \end{bmatrix}
    \begin{bmatrix}
        x_1 &  jx_1 & x_2 &  jx_2 \\
        x_3 & -jx_3 & x_4 & -jx_4
    \end{bmatrix} 
    \begin{bmatrix}
        \Re(\tilde{g}_1)\\
        \Im(\tilde{g}_1)\\
        \Re(\tilde{g}_2)\\
        \Im(\tilde{g}_2)
    \end{bmatrix}, \label{eq:generalMatrixFormLambdaVec}
\end{align}
with unknown parameters $\tilde{g}_1$ and $\tilde{g}_2$. It will be shown that the scaling of $f_1$ and $f_2$ can be compensated by choosing $\tilde{g}_1$ and $\tilde{g}_2$ as scaled versions of $g_1$ and $g_2$ with appropriately chosen scaling factors. Rewriting \eqref{eq:generalMatrixForm} and \eqref{eq:generalMatrixFormLambdaVec} and setting them equal produces
\begin{align}
    h_{\text{IQ}, i} &= f_1^* x_1 g_1 + f_1^* x_2 g_2 + f_2^* x_3 g_1^* + f_2^* x_4 g_2^* \nonumber \\
    &= \lambda_1 f_1^* x_1 \tilde{g}_1 + \lambda_1 f_1^* x_2 \tilde{g}_2 + \lambda_2 f_2^* x_3 \tilde{g}_1^* + \lambda_2 f_2^* x_4 \tilde{g}_2^*
\end{align}
yielding 
\begin{align}
    \tilde{g}_1   &= g_1   \frac{1}{\lambda_1},~
    \tilde{g}_2   = g_2   \frac{1}{\lambda_1},~
    \tilde{g}_1^* = g_1^* \frac{1}{\lambda_2},~
    \tilde{g}_2^* = g_2^* \frac{1}{\lambda_2}.
\end{align}
This shows that equality holds if $\lambda_1$ equals $\lambda_2^*$, and proofs that the IQ imbalance parameters are ambiguous after the bilinear estimation process, however, the IQ imbalance afflicted channel $\mathbf{h}_{\text{IQ}}$ is not affected by this ambiguity since these scaling factors cancel. The scaled parameters in $\mathbf{f}$ and $\mathbf{g}$ are passed to the subsequently discussed compensation approach to mitigate the IQ imbalance effects in the \ac{RDM}. It will be shown, that the increased noise floor can be reduced drastically.

\section{Compensation Strategy\label{sec:Compensation}}

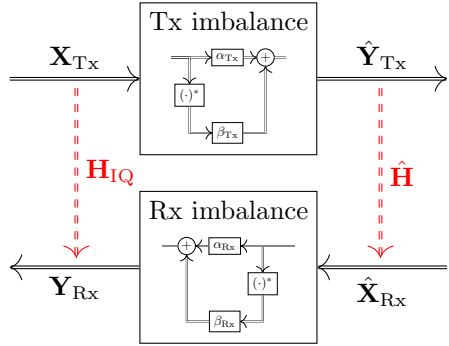
\begin{figure}
        \centering
        \begin{tikzpicture}[every node/.style=on grid]
            \node (center) [] at (0, 0) {};
        
            \node (inputTx) [above left=1.5 and 3 of center] {};
            \node (TxImbalance) [above left=1.5 and 0 of center, draw, align=center] 
                {
                    $\text{Tx imbalance}$ \\ 
                    \begin{tikzpicture}
                        \clip (-2.75,-0.75) rectangle (-1.25,0.75);
                        \node at (0,0) {\scalebox{0.5}{\input{figures/IQimbalanceBlockDiagram.tex}}};
                    \end{tikzpicture}
                };
            \node (outputTx) [above right=1.5 and 3 of center] {};
            \node (inputRx) [below right=1 and 3 of center] {};
            \node (RxImbalance) [below left=1 and 0 of center, draw, align=center] 
            {
                $\text{Rx imbalance}$ \\ 
                \begin{tikzpicture}
                    \clip (0.76,-0.75) rectangle (2.5,0.75);
                    \node at (0,0) {\scalebox{0.5}{\input{figures/IQimbalanceBlockDiagram_flip.tex}}};
                \end{tikzpicture}
            };
            \node (outputRx) [below left=1 and 3 of center] {};
            \node (inputTx_) [below left=-1.5 and 3 of center] {};

            \draw [->, double] (inputTx_) -- (TxImbalance) node [midway, above] {$\mathbf{X}_{\text{Tx}}$};
            \draw [->, double] (TxImbalance) -- (outputTx) node [midway, above] {$\hat{\mathbf{Y}}_{\text{Tx}}$};
            \draw [->, double] (inputRx) -- (RxImbalance) node [midway, below] {$\hat{\mathbf{X}}_{\text{Rx}}$};
            \draw [->, double] (RxImbalance) -- (outputRx) node [midway, below] {$\mathbf{Y}_{\text{Rx}}$};
            
            \node (inputH_Tx) [above left=1.5 and 2 of center] {};
            \node (outputH_Rx) [below left=1 and 2 of center] {};
            \draw[->, red, dashed, double] (inputH_Tx) -- (outputH_Rx) node[midway, right] {$\mathbf{H}_{\text{IQ}}$};
        
            \node (outputH_Tx) [above right=1.5 and 2 of center] {};
            \node (inputH_Rx) [below right=1 and 2 of center] {};
            \draw[->, red, dashed, double] (outputH_Tx) -- (inputH_Rx) node[midway, right] {$\hat{\mathbf{H}}$};
        \end{tikzpicture}
        \caption{Simplified block diagram  of the signal chain for the IQ imbalance compensation strategy.\label{fig:CompensationBlockDiagramm}} 
    \end{figure}

Subsequently to the parameter estimation process from the previous section, the compensation strategy as in Fig.~\ref{fig:CompensationBlockDiagramm} is described in this section. 
Considering all $N_{\text{sym}}$ \ac{OFDM} symbols, the frequency-domain data matrix $\mathbf{X}_{\text{Tx}}$ is transformed into $\hat{\mathbf{Y}}_{\text{Tx}}$ using the estimated parameters for the \ac{Tx} IQ imbalance by 
\begin{align}
    \hat{\mathbf{Y}}_{\text{Tx}} &= \hat{\alpha}_{\text{Tx}} \mathbf{X}_{\text{Tx}} + \hat{\beta}_{\text{Tx}} \mathbf{\underline{X}}_{\text{Tx}}^*.
\end{align}
For the \ac{Rx} path, one obtains
\begin{align}
    \mathbf{Y}_{\text{Rx}} &= \hat{\alpha}_{\text{Rx}} \hat{\mathbf{X}}_{\text{Rx}} + \hat{\beta}_{\text{Rx}} \hat{\mathbf{\underline{X}}}_{\text{Rx}}^*\label{eq:y_Rx},
\end{align}
and its complex conjugated and mirrored version can be calculated by
\begin{align}
    \mathbf{\underline{Y}}_{\text{Rx}}^* &= \hat{\alpha}_{\text{Rx}}^* \hat{\mathbf{\underline{X}}}_{\text{Rx}}^* + \hat{\beta}_{\text{Rx}}^* \hat{\mathbf{X}}_{\text{Rx}}\label{eq:y_Rx*}.
\end{align}
A combination of both yields the estimated frequency-domain data $\hat{\mathbf{X}}_{\text{Rx}}$ (for $|\hat{\alpha}_{\text{Rx}}|^2~-~|\hat{\beta}_{\text{Rx}}|^2~\neq~0$)
\begin{align}
    \hat{\mathbf{X}}_{\text{Rx}} &= \frac{1}{|\hat{\alpha}_{\text{Rx}}|^2 - | \hat{\beta}_{\text{Rx}}|^2} \left( \hat{\alpha}_{\text{Rx}}^* \mathbf{Y}_{\text{Rx}} -  \hat{\beta}_{\text{Rx}} \mathbf{\underline{Y}}_{\text{Rx}}^* \right).
\end{align}
The IQ imbalance compensated channel $\hat{\mathbf{H}}$ is subsequently calculated via 
\begin{align}
    \hat{\mathbf{H}} = \hat{\mathbf{X}}_{\text{Rx}} \oslash \hat{\mathbf{Y}}_{\text{Tx}}.\label{eq:IQcompensation}
\end{align}

\begin{figure}
        \centering
        \includegraphics[width=0.49\textwidth]{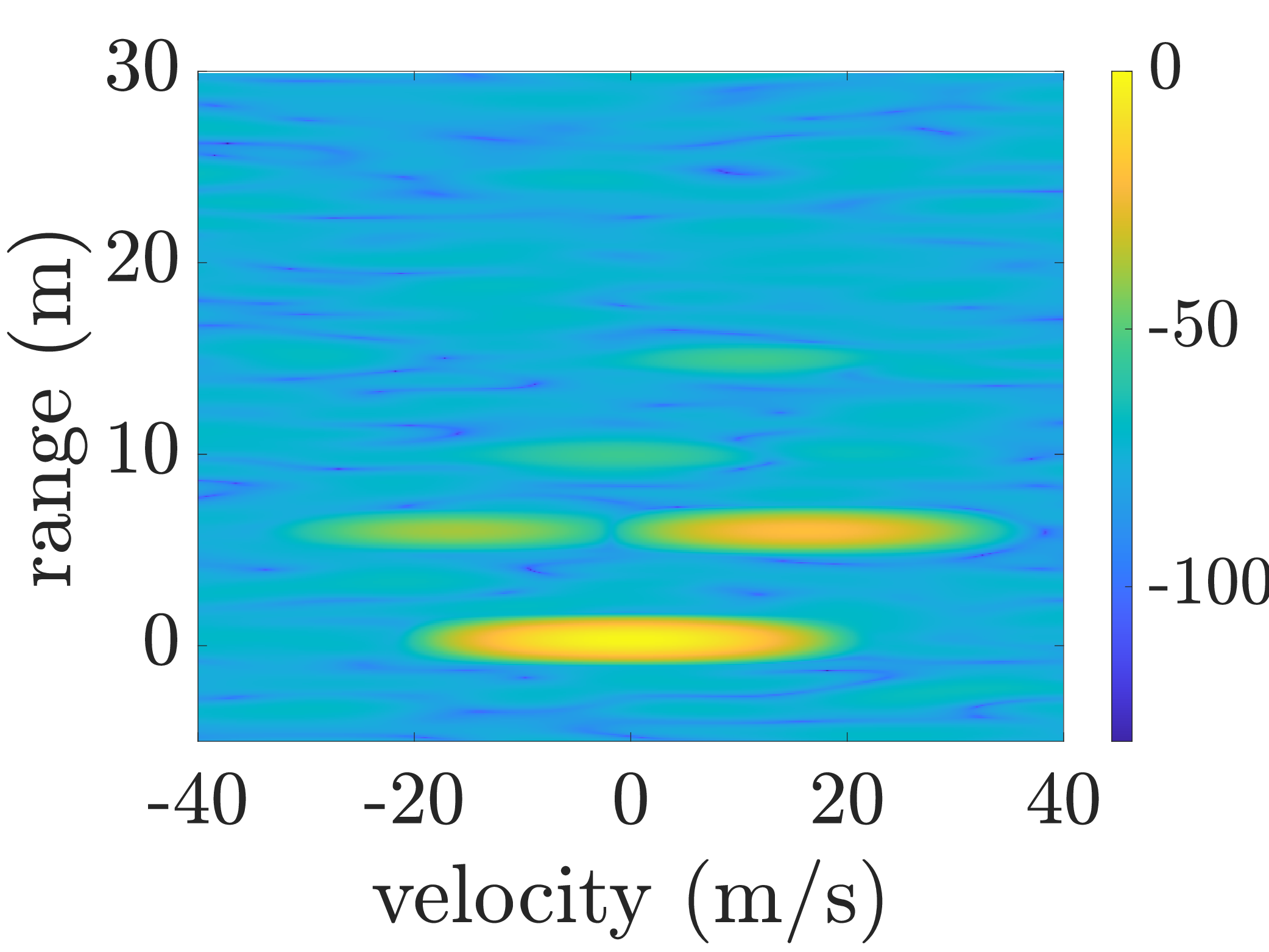}
        \caption{RDM of the compensated channel using the mixed complex-/real-valued bilinear LMS filter approach. The parameter estimation process is based on the estimated channel.}
        \label{fig:RDMcompIQ}
\end{figure}

In Fig.~\ref{fig:RDMcompIQ}, the \ac{RDM} of the IQ imbalance compensated channel is illustrated. For that, the mixed complex-/real-valued bilinear \ac{LMS} filter was used exemplarily to estimate parameters for the IQ imbalance compensation. This estimation process is based on the two most dominant peaks in the IQ imbalance afflicted channel. Compared to Fig.~\ref{fig:IQeffectsRDM}b, the noise floor is drastically reduced, however, the ghost object is still recognizable. As argued in \cite{LangWaveformsJCAS}, these ghost objects might be identified and cancelled by tracking algorithms.

\section{Simulations\label{sec:Simulations}}
In this work, different scenarios are simulated and analyzed. All simulations are performed with a 5G \ac{NR}-compliant \ac{OFDM} system. For that, the Matlab 5G toolbox is used to generate the \ac{Tx} \ac{PUSCH} data.

\subsection{Simulation Setup \label{sec:SimulationSetup}}
For the \ac{OFDM} system, the so-called numerology is chosen with $\mu = 3$, which yields a subcarrier spacing of $\Delta f = 2^{\mu} \cdot \SI{15}{kHz} = \SI{120}{kHz}$ \cite{3gpp.38.211}. The resource grid is assumed to be fully allocated and $N_{\text{sc}} = 3300$ subcarriers are occupied per \ac{OFDM} symbol. The \ac{Tx} resource grid $\mathbf{X}_{\text{Tx}}$ is filled with random 256 \ac{QAM} data symbols.

\begin{table}[!t]
    \centering
    \caption{Simulation setup: Leakage and objects in the surrounding of the \ac{UE}}
    \begin{tabular}{lrrr}
    \hline
        \toprule
        & \makecell{\textbf{normalized} \\ \textbf{power level}} & \multicolumn{1}{c}{\textbf{distance}} & \multicolumn{1}{c}{\textbf{velocity}} \\
        \midrule
        leakage  &   \SI{0}{dB} & \SI{0.5}{m} &   \SI{0}{m/s} \\ 
        \midrule
        object 1 & \SI{-40}{dB} &   \SI{6}{m} &  \SI{15}{m/s} \\
        \midrule
        object 2 & \SI{-43}{dB} &  \SI{15}{m} &  \SI{10}{m/s} \\
        \midrule
        object 3 & \SI{-47}{dB} &  \SI{10}{m} & \SI{0}{m/s} \\
        \bottomrule
    \end{tabular}
    \label{tab:simulationSetupScenario}
\end{table}

The scenario contains a strong leakage and 3 additional objects (detailed scenario setting given in Table~\ref{tab:simulationSetupScenario}). Two different sets of IQ imbalance parameters are considered in this work. The first set is based on a set used in \cite{schweizer2020iq,LangWaveformsJCAS} and is given by 
\begin{itemize}
    \item \textbf{FID transmitter IQ imbalance (literature-based):} amplitude error $\epsilon_{\text{Tx}} = 0.3$, and phase error $\Delta \phi = -20^{\circ}$, which correspond to imbalance parameters $\alpha_{\text{Tx}} = 0.9848 + j 0.0531$, and $\beta_{\text{Tx}} = 0.2954 + j 0.1736$,
    \item \textbf{FID receiver IQ imbalance (literature-based):} amplitude error $\epsilon_{\text{Tx}} = -0.1$, and phase error $\Delta \phi = -30^{\circ}$, which correspond to imbalance parameters $\alpha_{\text{Rx}} = 0.9659 + j 0.0259$, and $\beta_{\text{Rx}} = -0.0966 + j 0.2588$.
\end{itemize}
The corresponding \acp{EVM} for this literature-based set of IQ imbalance parameters are $\text{EVM}_{\text{Tx}} = 34.7\%$ and $\text{EVM}_{\text{Rx}} = 27.96\%$. According to \ac{3GPP} \cite[Table~6.4.2.1-1: Minimum requirements for \ac{EVM}]{3gppConfig.38.101-2}, the maximum \ac{EVM} level for 256 \ac{QAM} is given with \SI{3.5}{\%}, which serves as an upper bound in this work and the second set of IQ imbalance parameters is therefore defined with 
\begin{itemize}
    \item \textbf{FID transmitter IQ imbalance (3GPP-based):} amplitude error $\epsilon_{\text{Tx}} = -0.02$, and phase error $\Delta \phi = 2^{\circ}$, which correspond to the imbalance parameters $\alpha_{\text{Tx}} = 0.99985 - j 0.00035$, and $\beta_{\text{Tx}} = -0.0199 - j 0.0175$,
    \item \textbf{FID receiver IQ imbalance (3GPP-based):} amplitude error $\epsilon_{\text{Tx}} = -0.03$, and phase error $\Delta \phi = -2^{\circ}$, which correspond to the imbalance parameters $\alpha_{\text{Rx}} = 0.99985 + j 0.00052$, and $\beta_{\text{Rx}} = -0.0299 + j 0.01745$.
\end{itemize}
The resulting \ac{EVM} values are $\text{EVM}_{\text{Tx}} = 2.65\%$ and $\text{EVM}_{\text{Rx}} = 3.47\%$.

The parametrization for the mixed complex-/real-valued bilinear filters derived in \cite{MSc_Thesis_BP} are described in the following. The step size of the \ac{IWF} is set to $\mu=0.1$, and the number of used subcarriers to estimate the statistics $\mathbf{R}_{\mathbf{xx}}$ and $\mathbf{R}_{\mathbf{Xy}}$ required by the \ac{IWF} and \ac{AWF} is $N=3300$, which corresponds to one \ac{OFDM} symbol. For the  filter a step size of $\mu=0.02$, and for the \ac{RLS} $\lambda=0.95$ is used. Furthermore, for the \ac{RLS} estimator, the matrices $\mathbf{P}_{\text{h},0}$, and $\mathbf{P}_{\text{g},0}$ are initialized with $\mathbf{P}_{\text{h},0}=10^3~\mathbf{I}_{L}$, and $\mathbf{P}_{\text{g},0}=10^3~\mathbf{I}_{M}$, respectively. The normalization parameters for the  filter are set to $\alpha_{\text{h}}=0.7$, $\alpha_{\text{g}}=0.1$, $\delta_{\text{h}}=10^{-3}$, and $\delta_{\text{g}}=10^{-3}$. The mixed complex-/real-valued bilinear filter approaches are initialized with $\hat{\mathbf{f}}_0 = [1, 0]^{\text{T}}$, and $\hat{\mathbf{g}}_0 = [1,0, 0, 0]^{\text{T}}$, which are the IQ imbalance parameters without IQ imbalance.

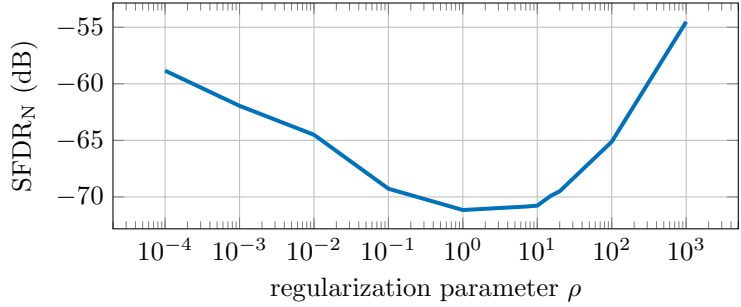
\begin{figure}
        \centering
        \input{figures/SFDR_RegularizationParam_overLambda}
        \caption{The SFDR$_{\text{N}}$ is illustrated for different regularization parameters $\rho$. The regularization parameter $\rho$ = 1 yields the best SFDR$_{\text{N}}$ performance.}
        \label{fig:SFDR_N_regPar}
\end{figure}

For the method proposed in \cite{ali2016ofdm}, one \ac{OFDM} symbol is used in time-domain (4096 data samples) to estimate the \ac{Tx} and \ac{Rx} IQ imbalance jointly with the channel. Simulations have shown, that the pseudo inverse in \cite[Eq.~(11)-(14)]{ali2016ofdm} leads to singularity issues. Hence, for the simulations in this paper, a regularization term $\rho$ is introduced with the regularization $\left( \mathbf{X}^T \mathbf{X} + \rho \mathbf{I}\right)^{-1} \mathbf{X}^T$. The regularization parameter $\rho$ is empirically evaluated. As we are interested to compensate the degrading effects of the IQ imbalance, the SFDR$_{\text{N}}$ is used as a performance measure (defined in the Section~\ref{sec:SFDR}). In Fig.~\ref{fig:SFDR_N_regPar}, the SFDR$_\text{N}$ is evaluated for different values of $\rho$. The results suggest that $\rho$~=~1 minimizes the SFDR$_\text{N}$ for our simulation scenario. After the iterative estimation process in \cite{ali2016ofdm}, the estimated parameters are used to calculate the channel matrix $\hat{\mathbf{H}}$ containing all \ac{OFDM} symbols. Therefore, \cite[Eq.~(11)]{ali2016ofdm} is used for each \ac{OFDM} symbol separately.

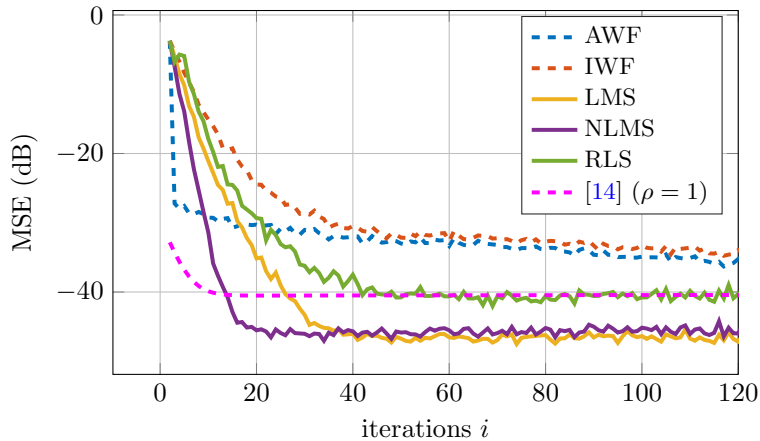
\begin{figure}
        \centering
        \input{figures/Error_80dB_lit}
        \caption{Estimation error plotted over the number of iterations averaged over 100 simulations.}
        \label{fig:errorPlot_literature}
\end{figure}

With that, the estimation errors corresponding to the literature-based IQ imbalance scenario (averaged over 100 simulations) are plotted over the iteration index in Fig.~\ref{fig:errorPlot_literature}. The estimation error in this plot is defined as the difference of the IQ imbalance afflicted channel element $\mathbf{h}_{\text{IQ},i}$ and the result of $\hat{\mathbf{f}}_i^{\text{T}} \mathbf{X}_i \hat{\mathbf{g}}_i$, with the iteration index $i$. The dashed curves represent the estimation error corresponding to non-sample adaptive methods (\ac{AWF}, \ac{IWF}, \cite{ali2016ofdm}). Note, the method proposed in \cite{ali2016ofdm} performs the estimation process based on one \ac{OFDM} symbol. Hence, this method requires a small number of iterations, however, each iteration consists of a huge number of calculations. The sample adaptive methods are plotted solid. It can be observed, that the estimation error of the \ac{AWF} decreases drastically within the first iteration, but convergence slows down soon after. The \ac{LMS} filter and the \ac{NLMS} filter reach convergence after about 35 and 18 iterations, respectively. The bilinear \ac{RLS} filter uses a forgetting factor of $\lambda=0.95$, which is quite small for this type of algorithm. However, simulations have shown that the bilinear \ac{RLS} filter tends to become instable for higher forgetting values. Interestingly, although the estimation error curves for the Wiener filters indicates less compensation performance, in the subsequent section it can be observed, that these filters outperform competitive methods investigated within this work.

\subsection{SFDR Evaluation \label{sec:SFDR}}
In this work, the power level difference between the most dominant peak in the \ac{RDM} (typically leakage) and the most dominant peak of the noise is defined as the \ac{SFDR}$_{\text{N}}$, representing the compensation performance regarding the increased noise floor. A second performance measure indicating the power level difference between an object peak in the \ac{RDM} and the power level of the corresponding ghost object at the negative velocity is denoted as \ac{SFDR}$_{\text{G}}$. The \ac{SFDR} evaluations are evaluated for different \ac{SNR} values and for both IQ imbalance parameter sets. The simulations without IQ imbalance serve as the lower bound of the compensation process.

\begin{figure}
        \centering
        \input{figures/SFDR_N_combined_400}
        \caption{The SFDR$_{\text{N}}$ performance is illustrated over different \acs{SNR} values and averaged over 400 simulations for the bilinear approaches and over 10 simulations for the method proposed in \cite{ali2016ofdm} due to its high computational complexity.}
        \label{fig:SFDR_N}
\end{figure}
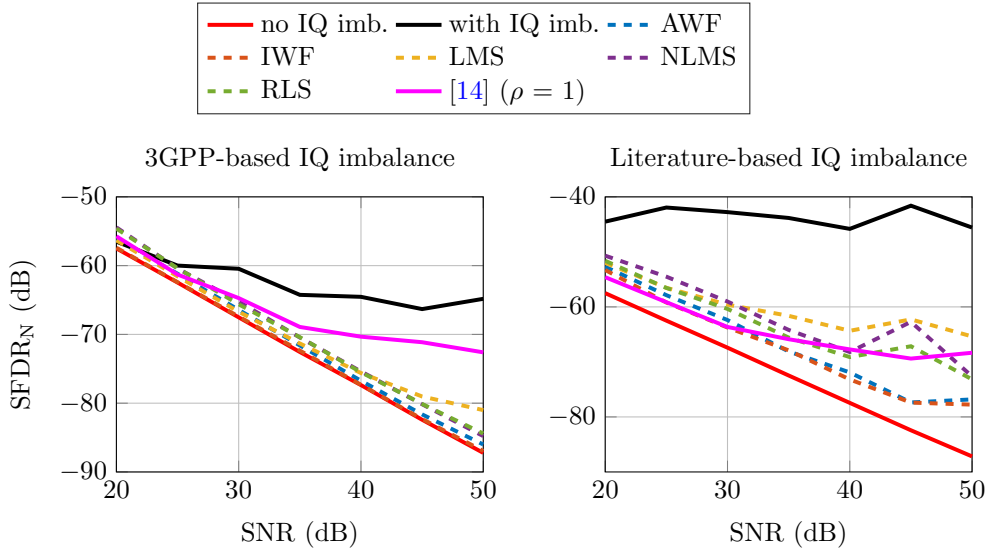

In Fig.~\ref{fig:SFDR_N}, the simulated \ac{SFDR}$_\text{N}$ values for the 3GPP-based, and the literature-based sets of IQ imbalance parameters are illustrated in the right, and in the left plot, respectively. The scenario without IQ imbalance is illustrated via the red curves, while the black solid line in these and the next plots show the IQ imbalance afflicted case without compensation. The performance of the aforementioned mixed complex-/real-valued bilinear filters as well as the comparison algorithm from literature \cite{ali2016ofdm} are illustrated. It can be observed, that the \ac{IWF} and \ac{AWF} outperform the other algorithms proposed in \cite{MSc_Thesis_BP} as well as the comparison algorithm in \cite{ali2016ofdm}. Especially for the 3GPP-based IQ imbalance case, the method proposed in \cite{ali2016ofdm} is clearly outperformed for \ac{SNR} values greater than \SI{30}{dB}. 

\begin{figure}
    \input{figures/SFDR_G_combined_400}
    \caption{The SFDR$_{\text{G}}$ performance is illustrated over different \acs{SNR} values and averaged over 400 simulations for the bilinear approaches and over 10 simulations for the method proposed in \cite{ali2016ofdm} due to its high computational complexity.}
    \label{fig:SFDR_G}
\end{figure}
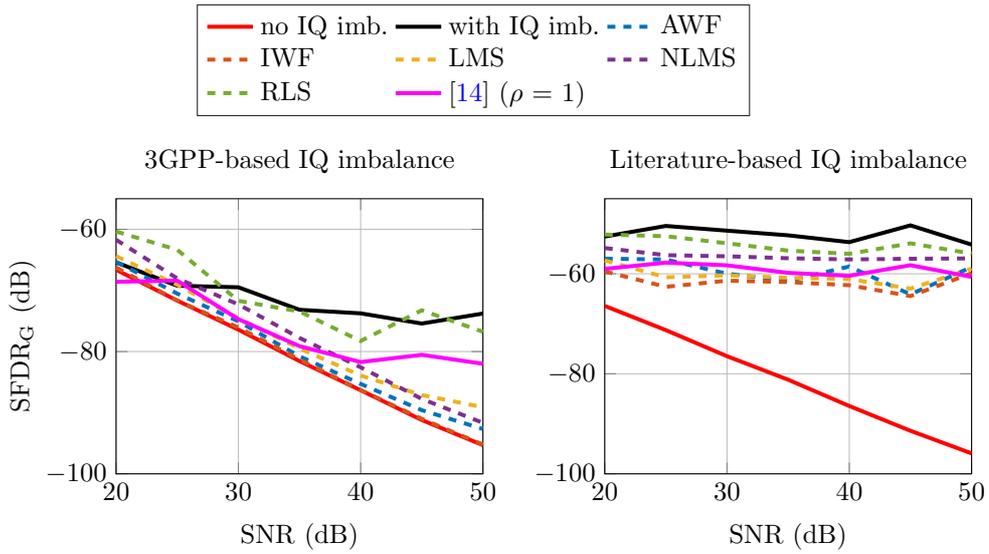

Fig.~\ref{fig:SFDR_G} shows the SFDR$_{\text{G}}$ performance, indicating that the \ac{IWF} and \ac{AWF} clearly outperform the other algorithms for the 3GPP-based IQ imbalance parameters. Different to that, for the literature-based IQ imbalance scenario basically all the methods are not able to decrease the power level of the ghost object with the proposed compensation procedure more than \SI{10}{dB}, with the \ac{IWF} and \ac{AWF} performing the best.

\subsection{MSE Evaluation}
The complex amplitudes of the objects in the \ac{RDM} is of importance, e.g., for digital beam forming in a \ac{MIMO} setting. Especially for, e.g., \ac{DOA} estimation, the reconstruction performance of the complex amplitudes is crucial. The true complex amplitudes are stored in a matrix $\mathbf{A} \in \mathbb{C}^{N_{\text{obj}} \times N_{\text{sim}}}$, where each row corresponds to an object, and each column corresponds to a simulation. The number of surrounding objects and the number of simulations is given with $N_{\text{obj}}$ and $ N_{\text{iter}}$, respectively. With the estimated complex amplitudes stored in the matrix $\hat{\mathbf{A}}$, the \ac{MSE} is evaluated by
\begin{align}
    \text{MSE} = \frac{1}{N_{\text{obj}} N_{\text{sim}}} \sum_{o=1}^{N_{\text{obj}}} \sum_{s=1}^{N_{\text{sim}}} |[\mathbf{A}]_{o,s} - [\hat{\mathbf{A}}]_{o,s}|^2.
\end{align}
Additionally, the averaged amplitude error $\epsilon$ and phase error $\Delta \phi$ is investigated via 
\begin{align}
    \epsilon = \frac{1}{N_{\text{obj}} N_{\text{sim}}} \sum_{o=1}^{N_{\text{obj}}} \sum_{s=1}^{N_{\text{sim}}} \frac{|[\mathbf{A}]_{o,s}| - |[\hat{\mathbf{A}}]_{o,s}|}{|[\mathbf{A}]_{o,s}|},
\end{align}
and
\begin{align}
    \Delta \phi = \frac{1}{N_{\text{obj}} N_{\text{sim}}} \sum_{o=1}^{N_{\text{obj}}} \sum_{s=1}^{N_{\text{sim}}} \left| \tan^{-1} \left(\frac{\Im ([\hat{\mathbf{A}}]_{o,s}/[\mathbf{A}]_{o,s})}{\Re([\hat{\mathbf{A}}]_{o,s}/[\mathbf{A}]_{o,s})}\right)\right|.
\end{align}
This is done for each discussed method, and the \ac{MSE} values corresponding to the 3GPP-based, and the literature-based IQ imbalance scenario are given in Table~\ref{tab:MSE_3GPP} and Table~\ref{tab:MSE_lit}, respectively. For that, 600 simulations and three different objects were evaluated (excluding the leakage). For each simulation run, the \ac{SNR} was randomly chosen from a uniform distribution between \SI{40}{dB} and \SI{50}{dB}. It can be observed, that the Wiener approaches clearly outperform the other implemented algorithms. With that, it can be summarized that the corresponding complex amplitude of an object in the surrounding can be accurately reconstructed after IQ imbalance compensation. Simulations have shown that this is the case for each single simulated object as well, however, the averaged values are given in this paper. This is explicitly the case for objects with a relative velocity of \SI{0}{m/s}, where the ghost object directly interferes with the true reflection. This might, e.g., provide the basis for \ac{DOA} implementations.

\begin{table}[!t]
    \centering
    \caption{Complex amplitude performance measures for the 3GPP-based set of IQ imbalance parameters.}
    \begin{tabular}{lccc}
        \hline
            \toprule
            & \textbf{MSE} & $\boldsymbol{\epsilon}$ & $\Delta \phi$ \\
            \midrule
            With IQ imb. & \SI{-62.5}{dB} & \SI{2.08}{\%} & 3.73$^{\circ}$  \\ 
            \midrule
            AWF & \SI{-77.8}{dB} & \SI{0.44}{\%} & 0.73$^{\circ}$ \\ 
            \midrule
            IWF & \SI{-78.2}{dB} & \SI{0.41}{\%} & 0.68$^{\circ}$ \\
            \midrule
            LMS & \SI{-61.6}{dB} & \SI{10.60}{\%} & 0.99$^{\circ}$ \\
            \midrule
            NLMS & \SI{-56.7}{dB} & \SI{19.02}{\%} & 0.95$^{\circ}$ \\ 
            \midrule
            RLS & \SI{-54.7}{dB} & \SI{18.01}{\%} & 1.89$^{\circ}$ \\ 
            \midrule
            \cite{ali2016ofdm} & \SI{-48.6}{dB} & \SI{8.02}{\%} & 11.47$^{\circ}$  \\ 
            \bottomrule
        \end{tabular}
    \label{tab:MSE_3GPP}
\end{table}
\begin{table}[!t]
    \centering
    \caption{Complex amplitude performance measures for the literature-based set of IQ imbalance parameters.}
    \begin{tabular}{lccc}
        \hline
            \toprule
            & \textbf{MSE} & $\boldsymbol{\epsilon}$ & $\Delta \phi$ \\
            \midrule
            With IQ imb. & \SI{-43.7}{dB} & \SI{18.86}{\%} & 36.92$^{\circ}$  \\ 
            \midrule
            AWF & \SI{-61.3}{dB} & \SI{5.87}{\%} & 5.43$^{\circ}$ \\ 
            \midrule
            IWF & \SI{-65.6}{dB} & \SI{5.33}{\%} & 3.75$^{\circ}$ \\
            \midrule
            LMS & \SI{-59.4}{dB} & \SI{8.74}{\%} & 3.55$^{\circ}$ \\
            \midrule
            NLMS & \SI{-53.4}{dB} & \SI{21.57}{\%} & 5.67$^{\circ}$ \\ 
            \midrule
            RLS & \SI{-53.2}{dB} & \SI{15.56}{\%} & 5.05$^{\circ}$ \\ 
            \midrule
            \cite{ali2016ofdm} & \SI{-46.9}{dB} & \SI{33.30}{\%} & 7.28$^{\circ}$  \\ 
            \bottomrule
        \end{tabular}
    \label{tab:MSE_lit}
\end{table}

\section{Conclusion\label{sec:Conclusion}}
This paper introduced a low-complexity method to compensate IQ imbalance effects for \ac{JCAS} systems. Unlike conventional approaches, which often struggle with leakage, our method exploits this effect as a source of information. The proposed approach is built on a mixed complex-/real-valued bilinear model, for which several applicable algorithms were investigated. The simulation results are based on the 5G \ac{NR} standard ensuring practical applicability in real-world applications. With these simulations, we demonstrated the performance in IQ imbalance compensation. Notably, the \ac{MSE} evaluation highlights a precise spectral peak reconstruction, while the \ac{SFDR} evaluation illustrates the noise floor reduction. Especially the Wiener approaches outperform competitive methods and provide efficient and robust solutions for IQ imbalance compensation.

\section*{Abbreviations}
\begin{acronym}
        \acro{3GPP}{3rd generation partnership project}
        \acro{AWF}{alternating Wiener filter}
        \acro{AWGN}{additive white Gaussian noise}
        \acro{BER}{bit error ratio}
        \acro{BS}{base station}
        \acro{CIR}{channel impulse response}
        \acro{CFAR}{constant false alarm rate}
        \acro{CP}{cyclic prefix}
        \acro{DOA}{direction of arrival}
        \acro{EVM}{error vector magnitude}
        \acro{FFT}{fast Fourier transform}
        \acro{FID}{frequency independent}
        \acro{FMCW}{frequency modulated continuous wave}
        \acro{FR}{frequency range}
        \acro{I}{in-phase}
        \acro{ICI}{inter-carrier interference}
        \acro{IFFT}{inverse fast Fourier transform}
        \acro{ISR}{image suppression ratio}
        \acro{IWF}{iterative Wiener filter}
        \acro{JCAS}{joint communications and sensing}
        \acro{LMS}{least mean squares}
        \acro{MIMO}{multiple input multiple output }
        \acro{MSE}{mean squared error}
        \acro{NLMS}{normalized \ac{LMS}}
        \acro{NR}{New Radio}
        \acro{OFDM}{orthogonal frequency-division multiplexing}
        \acro{Q}{quadrature-phase}
        \acro{QAM}{quadrature amplitude modulation}
        \acro{PUSCH}{physical uplink shared channel}
        \acro{RDM}{range-Doppler-map}
        \acro{RLS}{recursive least squares}
        \acro{Rx}{receive}
        \acro{SFDR}{spurious free dynamic range}
        \acro{SNR}{signal-to-noise ratio}
        \acro{Tx}{transmit}
        \acro{UE}{user equipment}
    \end{acronym}

    \section*{Availability of data and materials}
    Data sharing is not applicable to this article as no datasets were generated or analyzed during the current study.

    \section*{Competing interests}
    The authors declare that they have no competing interests
    
    \section*{Funding}
    The financial support by the Austrian Federal Ministry for Digital and Economic Affairs, the National Foundation for Research, Technology and Development and the Christian Doppler Research Association is gratefully acknowledged.
    
    \section*{Author contributions}
    All co-authors participated in the discussions to the content of this work and reviewed the manuscript.
    
    \section*{Acknowledgements}
    The financial support by the Austrian Federal Ministry for Digital and Economic Affairs, the National Foundation for Research, Technology and Development and the Christian Doppler Research Association is gratefully acknowledged.

    \bibliography{my-bibliography}

\end{document}

%% file: figures/IQimbalanceBlockDiagram.tex
\begin{tikzpicture}[every node/.style=on grid]
    \node (center) [] at (0, 0) {};

    \node (centerTx) [left=4 of center] {}; 
    \node (inputTx) [left=2.4 of centerTx] {};
    \node (betaTx) [below=2 of centerTx, draw] {$\beta_{\text{Tx}}$};
    \node (alphaTx) [right=0 of centerTx, draw] {$\alpha_{\text{Tx}}$};
    \node (conjTx) [above left=1 and 1 of betaTx, draw] {$(\cdot)^*$};
    \node (adderTx) [right=1 of centerTx, draw, circle, inner sep=1pt] {$+$};

    \node (channel) [left=0.5 of center, draw] {$h(t)$};
    \node (adder) [right=0.5 of center, draw, circle, inner sep=1pt] {$+$};
    \node (noise) [below right=1 and 0.5 of center] {};

    \node (centerRx) [right=3.5 of center] {}; 
    \node (betaRx) [below=2 of centerRx, draw] {$\beta_{\text{Rx}}$};
    \node (alphaRx) [right=0 of centerRx, draw] {$\alpha_{\text{Rx}}$};
    \node (conjRx) [above left=1 and 1 of betaRx, draw] {$(\cdot)^*$};
    \node (adderRx) [right=1 of centerRx, draw, circle, inner sep=1pt] {$+$};
    \node (outputRx) [right=1.5 of adderRx] {};

    \draw [->, double] (inputTx) -- (alphaTx) node [midway, above left] {$x_{\text{Tx}}(t)$};
    \draw [->, double] (alphaTx) -- (adderTx);
    \draw [->, double] (inputTx) -| (conjTx);
    \draw [->, double] (conjTx) |- (betaTx);
    \draw [->, double] (betaTx)  -| (adderTx);
    \draw [->, double] (adderTx) -- (channel) node [left=1.4, above] {$y_{\text{Tx}}(t)$};

    \draw [->, double] (channel) -- (adder);
    \draw [->, double] (noise) -- (adder) node [at start, below] {$n(t)$};

    \draw [->, double] (adder) -- (alphaRx) node [left=2, above] {$x_{\text{Rx}}(t)$};
    \draw [->, double] (alphaRx) -- (adderRx);
    \draw [->, double] (adder) -| (conjRx);
    \draw [->, double] (conjRx) |- (betaRx);
    \draw [->, double] (betaRx)  -| (adderRx);
    \draw [->, double] (adderRx) -- (outputRx) node [midway, above right] {$y_{\text{Rx}}(t)$};

\end{tikzpicture}

%% file: figures/IQimbalanceBlockDiagram_flip.tex
\begin{tikzpicture}[every node/.style=on grid]
    \node (center) [] at (0, 0) {};

    \node (centerTx) [left=4 of center] {}; 
    \node (inputTx) [left=2.4 of centerTx] {};
    \node (betaTx) [below=2 of centerTx, draw] {$\beta_{\text{Tx}}$};
    \node (alphaTx) [right=0 of centerTx, draw] {$\alpha_{\text{Tx}}$};
    \node (conjTx) [above left=1 and 1 of betaTx, draw] {$(\cdot)^*$};
    \node (adderTx) [right=1 of centerTx, draw, circle, inner sep=1pt] {$+$};

    \node (channel) [left=0.5 of center, draw] {$h(t)$};
    \node (adder) [right=0.5 of center, draw, circle, inner sep=1pt] {$+$};
    \node (noise) [below right=1 and 0.5 of center] {};

    \node (centerRx) [right=3.5 of center] {}; 
    \node (betaRx) [below=2 of centerRx, draw] {$\beta_{\text{Rx}}$};
    \node (alphaRx) [right=0 of centerRx, draw] {$\alpha_{\text{Rx}}$};
    \node (conjRx) [above right=1 and 1 of betaRx, draw] {$(\cdot)^*$};
    \node (adderRx) [left=1 of centerRx, draw, circle, inner sep=1pt] {$+$};
    \node (outputRx) [right=4.5 of adderRx] {};

    \draw [->, double] (inputTx) -- (alphaTx) node [midway, above left] {$x_{\text{Tx}}(t)$};
    \draw [->, double] (alphaTx) -- (adderTx);
    \draw [->, double] (inputTx) -| (conjTx);
    \draw [->, double] (conjTx) |- (betaTx);
    \draw [->, double] (betaTx)  -| (adderTx);
    \draw [->, double] (adderTx) -- (channel) node [left=1.4, above] {$y_{\text{Tx}}(t)$};

    \draw [->, double] (channel) -- (adder);
    \draw [->, double] (noise) -- (adder) node [at start, below] {$n(t)$};

    \draw [->, double] (adderRx) -- (adder) node [left=2, above] {$x_{\text{TRx}}(t)$};
    \draw [->, double] (alphaRx) -- (adderRx);
    \draw [->, double] (betaRx) -| (adderRx);
    \draw [->, double] (conjRx) |- (betaRx);
    \draw [->, double] (outputRx)  -| (conjRx);
    \draw [->, double] (outputRx) -- (alphaRx) node [midway, above right] {$y_{\text{TRx}}(t)$};

\end{tikzpicture}

%% file: figures/SFDR_RegularizationParam_overLambda
%
%
\definecolor{mycolor1}{rgb}{0.00000,0.44700,0.74100}%
\begin{tikzpicture}

\begin{semilogxaxis}[%
  compat=newest,
  width=.75\textwidth, 
  height = .35\textwidth, 
  grid, 
  xlabel={regularization parameter $\rho$},
  ylabel={SFDR$_{\text{N}}$ (dB)},
  legend pos=south east, 
  legend cell align=left,
  legend columns=3, 
  legend style={
    at={(0.05,1.2)}, 
    anchor=south west, 
    legend cell align=left, 
    align=left, 
    draw=white!15!black}
]
\addplot [color=mycolor1, forget plot, line width=1.5pt]
  table[row sep=crcr]{%
0.0001	-58.8404\\
0.001	-61.9482\\
0.01	-64.5062\\
0.1	-69.2738\\
1	-71.1554\\
7	-70.8513\\
10	-70.7702\\
15	-69.9182\\
20	-69.4993\\
100	-65.122\\
1000	-54.5243\\
};
\end{semilogxaxis}
\end{tikzpicture}%

%% file: figures/Error_80dB_lit
%
%
\definecolor{mycolor1}{rgb}{0.00000,0.44700,0.74100}%
\definecolor{mycolor2}{rgb}{0.85000,0.32500,0.09800}%
\definecolor{mycolor3}{rgb}{0.92900,0.69400,0.12500}%
\definecolor{mycolor4}{rgb}{0.49400,0.18400,0.55600}%
\definecolor{mycolor5}{rgb}{0.46600,0.67400,0.18800}%
\definecolor{mycolor6}{rgb}{1.00000,0.00000,1.00000}%
\begin{tikzpicture}

\begin{axis}[%
  compat=newest,
  width=.75\textwidth, 
  height = .49\textwidth, 
  grid, 
  xmax=120,
  xlabel={iterations $i$},
  ylabel={MSE (dB)},
  legend pos=south east, 
  legend cell align=left,
  legend columns=1, 
  legend style={
    at={(0.975, 0.985)}, 
    anchor=north east, 
    legend cell align=left, 
    align=left, 
    draw=white!15!black}
]
\addplot [color=mycolor1, dashed, line width=4.0pt, line width=1.5pt]
  table[row sep=crcr]{%
1	-inf\\
2	-3.73331144094291\\
3	-27.216708818088\\
4	-27.9906601813945\\
5	-27.5759960880379\\
6	-28.6293007065185\\
7	-28.7625365609752\\
8	-28.4344667067258\\
9	-28.7606027602921\\
10	-29.2548927532065\\
11	-29.2875099718751\\
12	-29.790598319445\\
13	-29.4331283299173\\
14	-29.3321370251324\\
15	-30.7035182636138\\
16	-30.5103605457237\\
17	-30.0765490556886\\
18	-30.2422095597424\\
19	-30.3426460168548\\
20	-30.2718544109229\\
21	-30.4921043758698\\
22	-30.3783968519841\\
23	-30.3162695699333\\
24	-31.6429063300782\\
25	-30.8456431783799\\
26	-31.2170052655563\\
27	-30.7450141021242\\
28	-31.5656823933559\\
29	-31.7058437142646\\
30	-31.0305008533878\\
31	-31.1857583627597\\
32	-30.9061967834842\\
33	-31.4974898677207\\
34	-31.8017420120631\\
35	-32.4047762757327\\
36	-31.9197582470328\\
37	-31.9308914452389\\
38	-32.1470922137143\\
39	-32.0932876527089\\
40	-31.9805355179932\\
41	-31.975730218026\\
42	-33.0011722699893\\
43	-32.6807397759647\\
44	-31.92527698627\\
45	-32.4049000567042\\
46	-32.8730021233285\\
47	-32.796859928637\\
48	-32.4738426320095\\
49	-32.6532260939626\\
50	-32.9970137269789\\
51	-33.1752547548105\\
52	-32.7018634670303\\
53	-32.8219775081313\\
54	-32.0906471805617\\
55	-32.9247926746976\\
56	-32.9300966396339\\
57	-32.4366313178777\\
58	-32.5636836424364\\
59	-32.5096536875239\\
60	-33.3590368940394\\
61	-32.2275674260522\\
62	-32.893819148519\\
63	-33.1963745871428\\
64	-33.3107078450195\\
65	-33.1962134282192\\
66	-32.6011792426558\\
67	-32.9260290787278\\
68	-33.065343141082\\
69	-33.1224990993875\\
70	-33.2069977341868\\
71	-33.5066706966304\\
72	-33.5081226071536\\
73	-33.7715286379351\\
74	-33.9570861186531\\
75	-33.6566138753635\\
76	-33.6557912069246\\
77	-33.6865180804568\\
78	-33.4196521277371\\
79	-33.7909340218732\\
80	-33.8816775222945\\
81	-33.614431437501\\
82	-33.6432631930349\\
83	-34.1255452024555\\
84	-34.1589153009638\\
85	-34.2808148922095\\
86	-34.7541334905849\\
87	-33.9192023307195\\
88	-33.697996584875\\
89	-33.9499383659306\\
90	-34.6310784504353\\
91	-34.6802077885259\\
92	-34.0011949713386\\
93	-34.6262708062276\\
94	-34.6535468661749\\
95	-34.9540509639289\\
96	-34.9336320786473\\
97	-34.9062401970765\\
98	-35.0439166822963\\
99	-35.0300972115867\\
100	-34.8845334052255\\
101	-35.0263412466339\\
102	-34.9850014470509\\
103	-35.0025598599266\\
104	-35.0454873778416\\
105	-34.5850605138965\\
106	-35.6301204534323\\
107	-34.9538525297361\\
108	-35.098947842283\\
109	-35.2136626739012\\
110	-35.0831318314386\\
111	-34.8484993370896\\
112	-35.6295810136361\\
113	-35.4941209568731\\
114	-35.6518048002084\\
115	-35.9392928241613\\
116	-35.6400694742722\\
117	-36.3157436236453\\
118	-36.0354659515146\\
119	-35.8037480265396\\
120	-35.2298364885366\\
121	-35.7854275926999\\
122	-35.6159336585613\\
123	-35.7849504542347\\
124	-36.1402060601392\\
125	-35.841444053889\\
126	-36.2123071251827\\
127	-35.7831232122601\\
128	-35.7358220530662\\
129	-35.3268930183698\\
130	-35.7697122720851\\
131	-36.1132317660309\\
132	-35.8488109314639\\
133	-35.432642228631\\
134	-36.2730295539407\\
135	-36.1385189549468\\
136	-35.9803337648723\\
137	-35.0729254999117\\
138	-36.3212772989604\\
139	-35.924790850473\\
140	-35.9991712469841\\
141	-36.135255847637\\
142	-35.8087869196672\\
143	-35.5130819569325\\
144	-36.1002238229006\\
145	-36.1326026547285\\
146	-35.7513133865777\\
147	-35.626630339593\\
148	-35.6888389398815\\
149	-35.7445619149773\\
150	-35.5274566098311\\
};
\addlegendentry{\scriptsize AWF}

\addplot [color=mycolor2, dashed, line width=4.0pt, line width=1.5pt]
  table[row sep=crcr]{%
1	-inf\\
2	-3.73331144094291\\
3	-5.27593901816406\\
4	-6.73745514751965\\
5	-8.33029104132427\\
6	-9.72288923545299\\
7	-11.1227160811568\\
8	-12.9659398382474\\
9	-14.0308698357247\\
10	-15.0878709416716\\
11	-16.0559268961659\\
12	-17.2287092833746\\
13	-19.1753030820567\\
14	-19.4779637123302\\
15	-19.2626583657096\\
16	-20.8816539529655\\
17	-21.8716231422436\\
18	-22.4236411306616\\
19	-23.8452856419198\\
20	-24.4563003643478\\
21	-24.5210300586172\\
22	-25.3479664518657\\
23	-25.6616263361255\\
24	-26.205936752046\\
25	-26.3738499384961\\
26	-27.4263979382392\\
27	-27.3101051362456\\
28	-27.7932692293295\\
29	-28.7126129748696\\
30	-28.6097869163878\\
31	-28.809631810286\\
32	-28.2610353288636\\
33	-29.1012460375351\\
34	-29.4549621940079\\
35	-30.7130269208602\\
36	-30.362436666402\\
37	-30.1361664000332\\
38	-30.7567547899492\\
39	-30.8273200444467\\
40	-30.4755678805054\\
41	-30.7097741975389\\
42	-31.1892750903808\\
43	-31.2860329133629\\
44	-30.6231814410666\\
45	-31.2124639148556\\
46	-31.7539984526184\\
47	-32.2524508096186\\
48	-31.5272359966047\\
49	-31.7433633423322\\
50	-32.1537929983224\\
51	-32.1821379325915\\
52	-31.5783017853586\\
53	-31.9374700937316\\
54	-31.2923785434462\\
55	-31.7586839053766\\
56	-31.8454478204031\\
57	-31.6541696343457\\
58	-31.5483892019334\\
59	-31.4911669212354\\
60	-32.6622456033969\\
61	-31.1985702283964\\
62	-31.8721647601578\\
63	-32.3653411666192\\
64	-32.4331902745955\\
65	-32.1703094157743\\
66	-31.8307420778835\\
67	-31.9101289414733\\
68	-31.9341185389238\\
69	-32.285134634274\\
70	-32.2283770961877\\
71	-32.4494035813315\\
72	-32.3662416453486\\
73	-32.622330423137\\
74	-32.9567104734395\\
75	-32.5097844677469\\
76	-32.3424555293806\\
77	-32.8145457774949\\
78	-32.4027900059897\\
79	-32.9459304651854\\
80	-32.7266801362337\\
81	-32.6141101280596\\
82	-32.6650346421962\\
83	-32.7436844644859\\
84	-33.1647711660328\\
85	-33.0326617940444\\
86	-33.2698215126916\\
87	-32.721264910116\\
88	-32.9069157508798\\
89	-32.6398816574509\\
90	-33.2927282234697\\
91	-33.9455266064278\\
92	-33.0279053314503\\
93	-33.3509594965488\\
94	-33.3495711133367\\
95	-33.8385579353207\\
96	-33.5340391120396\\
97	-33.5568862401211\\
98	-33.7982642216876\\
99	-33.6301827664969\\
100	-33.1932214826156\\
101	-33.6101605075971\\
102	-33.6983630583529\\
103	-33.6654460085331\\
104	-34.1641114585085\\
105	-33.3596898536727\\
106	-34.5630940634684\\
107	-33.5424154033376\\
108	-33.7745202329501\\
109	-33.6949906299406\\
110	-33.5932288631312\\
111	-33.4372808294147\\
112	-34.0226302186277\\
113	-33.7516999895583\\
114	-34.4177667737245\\
115	-34.4574386801978\\
116	-34.1870752390022\\
117	-34.7757592901143\\
118	-34.6018537328356\\
119	-34.4888953494748\\
120	-33.933895540415\\
121	-34.3432616170562\\
122	-34.6264410502175\\
123	-34.6694793516594\\
124	-34.1861551534648\\
125	-34.2608878128398\\
126	-34.869149268024\\
127	-34.4071275107323\\
128	-34.3398815004301\\
129	-33.8672434834351\\
130	-34.8735680407577\\
131	-34.551526442296\\
132	-34.0545779591704\\
133	-34.2281056988677\\
134	-34.9256286227442\\
135	-34.7758477252218\\
136	-34.672123832734\\
137	-34.2211997494571\\
138	-34.6223820807324\\
139	-34.3253523663007\\
140	-34.8093019734443\\
141	-34.523290125477\\
142	-34.5316517399227\\
143	-34.6812543438553\\
144	-34.6419838559404\\
145	-34.8279885863538\\
146	-34.5087140406106\\
147	-34.6942575686228\\
148	-34.6355073533899\\
149	-34.8090282141806\\
150	-33.8743554572253\\
};
\addlegendentry{\scriptsize IWF}

\addplot [color=mycolor3, line width=4.0pt, line width=1.5pt]
  table[row sep=crcr]{%
1	-inf\\
2	-3.73331144094291\\
3	-6.00497703583627\\
4	-8.5313886478825\\
5	-10.141512962026\\
6	-13.1638921343922\\
7	-14.6816913146218\\
8	-17.6845101526607\\
9	-19.2751258294118\\
10	-21.0987726164016\\
11	-22.7929667310659\\
12	-24.5448773330181\\
13	-25.2605680041707\\
14	-27.2577591741353\\
15	-27.1506366567435\\
16	-29.7404072643907\\
17	-30.6814787715733\\
18	-32.2064115275385\\
19	-33.2309189656716\\
20	-35.1443946815051\\
21	-35.0182634796069\\
22	-36.9973735230398\\
23	-38.1265473421851\\
24	-37.1970171484637\\
25	-39.0204661293517\\
26	-40.2933065336347\\
27	-40.9423220809404\\
28	-41.3220258120454\\
29	-42.4050028085984\\
30	-43.2972372138022\\
31	-44.7525543391303\\
32	-44.4703880655997\\
33	-44.891919393263\\
34	-45.1791707148137\\
35	-45.2319149108445\\
36	-45.6146231459387\\
37	-45.3720333593156\\
38	-46.1640268138433\\
39	-46.1138335424421\\
40	-46.0255410027057\\
41	-46.9152031966194\\
42	-46.575287874376\\
43	-46.0390120054526\\
44	-46.3789378561831\\
45	-46.2862488059657\\
46	-46.6183743742071\\
47	-46.7477435495474\\
48	-47.2925470608762\\
49	-46.4747171827649\\
50	-46.731739774979\\
51	-47.0664799841805\\
52	-47.1581687918408\\
53	-46.6159316153536\\
54	-47.0404155942234\\
55	-46.6159265168628\\
56	-46.7066961772854\\
57	-46.3782800136635\\
58	-46.0667204534298\\
59	-46.0760030286544\\
60	-46.5474281816426\\
61	-46.6331411962194\\
62	-46.4121838968937\\
63	-46.5315587066626\\
64	-46.6576158443771\\
65	-47.1593514676155\\
66	-47.0093257296373\\
67	-46.7517438183023\\
68	-46.7715372843242\\
69	-47.3023963094483\\
70	-46.9462929335132\\
71	-46.1506504895317\\
72	-46.4109392077712\\
73	-46.1342583526975\\
74	-47.5142790952254\\
75	-46.6035154456627\\
76	-46.4855732393141\\
77	-46.3991476622003\\
78	-46.3378792617662\\
79	-46.2773264384549\\
80	-46.2235960797745\\
81	-46.8684191119082\\
82	-47.4518951914574\\
83	-46.7089340966072\\
84	-46.6824929199475\\
85	-46.143468868638\\
86	-46.6778937492086\\
87	-46.7335437728177\\
88	-46.3999258232518\\
89	-45.8324140804808\\
90	-46.5641052337734\\
91	-46.159867176456\\
92	-46.3996368376165\\
93	-46.291871623291\\
94	-46.3678338067944\\
95	-47.3530285541462\\
96	-46.3924540473544\\
97	-45.6472309304045\\
98	-46.2459069671586\\
99	-46.5220441997671\\
100	-46.3590362849025\\
101	-46.7638287042759\\
102	-46.3986754040816\\
103	-46.7845505137991\\
104	-47.1653721140083\\
105	-46.1141315695595\\
106	-46.4083227943279\\
107	-46.6872800709011\\
108	-46.554518082969\\
109	-45.4260678650943\\
110	-46.7269707469384\\
111	-47.0373391495211\\
112	-46.5483774255398\\
113	-46.9255117319071\\
114	-47.0571128640408\\
115	-47.3511592813549\\
116	-46.1679074474768\\
117	-46.9720096828417\\
118	-45.8344873739321\\
119	-46.791108623135\\
120	-47.160797982163\\
121	-46.176858352616\\
122	-46.4177872201212\\
123	-46.5092691808009\\
124	-46.6623048198932\\
125	-45.7303258042824\\
126	-46.4244722883157\\
127	-46.9188548028536\\
128	-46.7124897561008\\
129	-45.6522058704219\\
130	-45.6366150575302\\
131	-46.135273687495\\
132	-46.4910459997964\\
133	-46.8492166233653\\
134	-46.170709929054\\
135	-46.3329001565152\\
136	-46.175138923539\\
137	-46.5601549560478\\
138	-46.1458320749713\\
139	-46.6487088236815\\
140	-46.6529425879801\\
141	-47.5270528146798\\
142	-46.1224926428308\\
143	-46.4359037202849\\
144	-46.342660990302\\
145	-46.0858408718282\\
146	-47.2106132502979\\
147	-46.9812461613188\\
148	-47.2249966714523\\
149	-47.5184893677427\\
150	-46.6357429872739\\
};
\addlegendentry{\scriptsize LMS}

\addplot [color=mycolor4, line width=4.0pt, line width=1.5pt]
  table[row sep=crcr]{%
1	-inf\\
2	-3.73331144094291\\
3	-7.44138098891066\\
4	-11.3830475512196\\
5	-13.9312402763495\\
6	-18.3143417311598\\
7	-22.2407009732706\\
8	-25.3409307666956\\
9	-28.4407130600673\\
10	-31.4010772964792\\
11	-35.8706677531996\\
12	-37.6517075257608\\
13	-39.2390008197352\\
14	-40.7489838691744\\
15	-42.8627871282036\\
16	-44.431048362256\\
17	-44.3352101387819\\
18	-45.1313938638855\\
19	-45.0371100011276\\
20	-45.5315086735492\\
21	-45.3846434697848\\
22	-45.7055766693857\\
23	-46.5452176161386\\
24	-45.4011598000298\\
25	-45.9176020735507\\
26	-46.2426199578997\\
27	-45.3596522801853\\
28	-45.9352141216193\\
29	-45.9827948425377\\
30	-46.0336165413215\\
31	-46.3789516358474\\
32	-45.9955505215465\\
33	-46.0585404465437\\
34	-47.0382342360257\\
35	-45.8758361647572\\
36	-46.2274610217192\\
37	-45.2589868086083\\
38	-45.8370266540382\\
39	-45.4969591765148\\
40	-45.6693954744326\\
41	-46.3554258133749\\
42	-45.8548914091426\\
43	-45.748638401376\\
44	-46.2980671693692\\
45	-45.8440525773033\\
46	-45.2781639294714\\
47	-45.9927211232078\\
48	-46.7681878269165\\
49	-45.6990020544656\\
50	-45.8519857122406\\
51	-45.8019068897422\\
52	-46.0930213957259\\
53	-46.1184863709117\\
54	-46.4256600509377\\
55	-45.7976723399369\\
56	-45.2788990115734\\
57	-45.5871843838491\\
58	-45.3016862653382\\
59	-45.2247676420787\\
60	-45.5219025378019\\
61	-46.1034699604704\\
62	-45.9688836508007\\
63	-45.3214444895461\\
64	-45.2232884839482\\
65	-45.9985298549417\\
66	-46.442963007542\\
67	-45.8561670093875\\
68	-45.2628460148304\\
69	-46.1264705194721\\
70	-45.9340736589727\\
71	-45.5926479319069\\
72	-46.0966033586176\\
73	-45.0382948518332\\
74	-46.1061910038476\\
75	-45.7809518730679\\
76	-45.9171095612638\\
77	-45.1461781839003\\
78	-46.1319988486375\\
79	-45.7208466874956\\
80	-45.1908672597904\\
81	-45.4272754661896\\
82	-46.5320506107768\\
83	-45.5041657688176\\
84	-45.6502206675209\\
85	-45.4735840880071\\
86	-46.5159520484658\\
87	-45.8442223786038\\
88	-45.7110071888298\\
89	-44.723907538233\\
90	-45.0040863928408\\
91	-45.5123897713502\\
92	-45.7310633775999\\
93	-44.8607693616939\\
94	-45.2923483815008\\
95	-45.8477394010114\\
96	-45.65673173033\\
97	-45.0928327073115\\
98	-45.7363757944351\\
99	-45.6485664290263\\
100	-45.1484876034078\\
101	-45.2794936829322\\
102	-45.5043534812622\\
103	-45.0367876778554\\
104	-46.0849167691308\\
105	-45.1063773987949\\
106	-45.7321146057328\\
107	-46.0730934771009\\
108	-45.1167523777173\\
109	-44.8271368847956\\
110	-46.0560565068825\\
111	-45.4179230965631\\
112	-45.2647701999898\\
113	-45.9756852277897\\
114	-46.2069796851018\\
115	-45.904424320594\\
116	-44.2212765677962\\
117	-46.4088532820884\\
118	-45.156621714587\\
119	-45.9621506540039\\
120	-45.9903399959971\\
121	-44.6423861738786\\
122	-45.0479667650138\\
123	-46.0445488627861\\
124	-45.6855049988083\\
125	-44.6294809401526\\
126	-45.7078820176426\\
127	-46.0815348409683\\
128	-46.0297414044604\\
129	-44.7356383204973\\
130	-44.4966066132291\\
131	-45.3740793077342\\
132	-46.0455577504653\\
133	-45.2629860377319\\
134	-45.3821386946487\\
135	-44.9236023489663\\
136	-45.2135870894831\\
137	-45.6512615934683\\
138	-44.837648529583\\
139	-45.3403540247952\\
140	-45.93028372638\\
141	-46.5249222637982\\
142	-45.3849815773242\\
143	-45.2025430022764\\
144	-45.4681525420489\\
145	-46.0016524449871\\
146	-45.9809243449717\\
147	-45.7775199671561\\
148	-46.274666108903\\
149	-46.2071154834086\\
150	-45.9779661314766\\
};
\addlegendentry{\scriptsize NLMS}

\addplot [color=mycolor5, line width=4.0pt, line width=1.5pt]
  table[row sep=crcr]{%
1	-inf\\
2	-3.73331144094291\\
3	-6.80557596200973\\
4	-5.71943370114632\\
5	-5.9132125224939\\
6	-9.64167645283803\\
7	-11.4072922091082\\
8	-13.9546022770599\\
9	-15.3401465541585\\
10	-17.9008729335936\\
11	-19.7685174751534\\
12	-21.937665496361\\
13	-21.8012607404128\\
14	-24.4657758587441\\
15	-24.5357053191949\\
16	-25.3456527253114\\
17	-27.5602011686271\\
18	-28.0843377952227\\
19	-28.6373990771658\\
20	-29.3267173499375\\
21	-29.4261679839912\\
22	-32.5436503581992\\
23	-31.0231203266065\\
24	-31.7662949399702\\
25	-33.0771068962152\\
26	-33.5441783262315\\
27	-33.3847839451611\\
28	-34.1629782811394\\
29	-36.0292566815167\\
30	-36.2093913994657\\
31	-37.1407183941545\\
32	-36.5159887221176\\
33	-37.2569312360764\\
34	-38.5413905360436\\
35	-37.9413631497367\\
36	-37.0020547377633\\
37	-37.612036984773\\
38	-39.2187411742112\\
39	-38.4216924244104\\
40	-37.4933493228093\\
41	-38.8381223568357\\
42	-40.4410984559345\\
43	-39.5930157277153\\
44	-39.3408476446441\\
45	-40.233683768867\\
46	-40.3540737535981\\
47	-40.3931158705769\\
48	-39.9516601499221\\
49	-40.7956913641117\\
50	-40.7742790596415\\
51	-40.3133873188644\\
52	-40.21472407227\\
53	-40.2588391393651\\
54	-39.2922615150542\\
55	-40.5658622689722\\
56	-40.5723896669582\\
57	-40.4648538911792\\
58	-40.2364380699929\\
59	-40.6208369024038\\
60	-40.6321601110941\\
61	-40.4041564152099\\
62	-41.0351137880567\\
63	-39.9798151637296\\
64	-40.9234609056427\\
65	-40.9418157331003\\
66	-40.7529643292771\\
67	-41.8827152971828\\
68	-40.5059461533809\\
69	-42.0178117836616\\
70	-40.6102477137612\\
71	-40.9015068773233\\
72	-40.0078583836464\\
73	-40.5270558140761\\
74	-41.2381729535393\\
75	-40.970347672721\\
76	-40.8363459488907\\
77	-41.358610584421\\
78	-41.2457419203816\\
79	-41.4750999938647\\
80	-40.5778181501299\\
81	-41.0996873428922\\
82	-41.1649732781186\\
83	-40.506500561315\\
84	-40.2861658641916\\
85	-40.6140039516904\\
86	-41.6620994544928\\
87	-39.7731307558546\\
88	-40.3620256518328\\
89	-40.4056346342106\\
90	-40.5782343412138\\
91	-40.5508812291118\\
92	-40.2751446676708\\
93	-40.1763188524331\\
94	-39.9952875391569\\
95	-40.7487043409391\\
96	-40.2230049670375\\
97	-40.596585773416\\
98	-40.1517006229068\\
99	-41.0307518505291\\
100	-40.7710861471695\\
101	-40.7796774275766\\
102	-40.2897714945901\\
103	-39.750456688853\\
104	-41.0482798678597\\
105	-39.5651216841603\\
106	-41.3167464102243\\
107	-40.7017174002366\\
108	-40.2232673003647\\
109	-40.7863749425224\\
110	-39.8281434864618\\
111	-40.5512619154411\\
112	-40.9422746180771\\
113	-40.0930829675442\\
114	-41.1285180297633\\
115	-41.406635088908\\
116	-39.7946533364333\\
117	-40.9440300615218\\
118	-38.9910060839632\\
119	-40.9018611059841\\
120	-40.2719317990488\\
121	-39.8748258690774\\
122	-40.1978760414312\\
123	-40.2231093212124\\
124	-40.063316288998\\
125	-40.2049284955293\\
126	-40.4409208406682\\
127	-39.8184756960222\\
128	-40.2818135874381\\
129	-39.1741613863095\\
130	-40.816671645348\\
131	-40.1851536062307\\
132	-39.9963558949346\\
133	-39.7411115348149\\
134	-39.7803574871755\\
135	-40.2455036904358\\
136	-39.9529881205882\\
137	-39.6591591802545\\
138	-39.7525705407959\\
139	-40.5293179648417\\
140	-40.8122560066153\\
141	-40.2602777797502\\
142	-39.0816068583283\\
143	-40.3882107644759\\
144	-40.0112374009431\\
145	-39.0958991764336\\
146	-40.7025977444891\\
147	-39.6150937855687\\
148	-40.4648641429839\\
149	-40.2725149925257\\
150	-39.8455177300815\\
};
\addlegendentry{\scriptsize RLS}

\addplot [color=mycolor6, dashed, line width=4.0pt, line width=1.5pt]
  table[row sep=crcr]{%
1	-inf\\
2	-32.8277901595534\\
3	-34.3044755682926\\
4	-35.6354067887913\\
5	-36.7925985906629\\
6	-37.7606763705807\\
7	-38.536709767677\\
8	-39.132294725174\\
9	-39.572842135385\\
10	-39.8876210395704\\
11	-40.1062210470951\\
12	-40.2540390782278\\
13	-40.3523068621129\\
14	-40.4166776012264\\
15	-40.4579042997909\\
16	-40.4837556371891\\
17	-40.4997052554021\\
18	-40.5092233837767\\
19	-40.5145514998784\\
20	-40.5172325093884\\
21	-40.5182226803096\\
22	-40.5181527919154\\
23	-40.5174404267794\\
24	-40.5163505052542\\
25	-40.5150469562949\\
26	-40.5136305673347\\
27	-40.5121666587358\\
28	-40.5106927084625\\
29	-40.5092297773718\\
30	-40.5077891648884\\
31	-40.506376225039\\
32	-40.50499277757\\
33	-40.5036386180397\\
34	-40.502312453037\\
35	-40.5010124774172\\
36	-40.4997367209523\\
37	-40.498483244213\\
38	-40.4972502407233\\
39	-40.4960360837615\\
40	-40.4948393405536\\
41	-40.4936587677197\\
42	-40.4924932968318\\
43	-40.4913420155721\\
44	-40.490204147747\\
45	-40.4890790339616\\
46	-40.4879661138765\\
47	-40.4868649104518\\
48	-40.4857750162488\\
49	-40.4846960817108\\
50	-40.483627805249\\
51	-40.4825699249234\\
52	-40.4815222115101\\
53	-40.4804844627514\\
54	-40.4794564986032\\
55	-40.4784381573231\\
56	-40.4774292922509\\
57	-40.4764297691715\\
58	-40.4754394641527\\
59	-40.4744582617768\\
60	-40.4734860536934\\
61	-40.4725227374388\\
62	-40.4715682154713\\
63	-40.4706223943814\\
64	-40.4696851842514\\
65	-40.4687564981262\\
66	-40.4678362515887\\
67	-40.4669243624064\\
68	-40.4660207502475\\
69	-40.4651253364439\\
70	-40.4642380438025\\
71	-40.4633587964467\\
72	-40.4624875196873\\
73	-40.4616241399175\\
74	-40.4607685845262\\
75	-40.4599207818245\\
76	-40.4590806609874\\
77	-40.4582481520043\\
78	-40.457423185638\\
79	-40.4566056933887\\
80	-40.4557956074676\\
81	-40.4549928607706\\
82	-40.4541973868566\\
83	-40.4534091199322\\
84	-40.4526279948336\\
85	-40.451853947015\\
86	-40.4510869125344\\
87	-40.4503268280461\\
88	-40.4495736307875\\
89	-40.4488272585744\\
90	-40.4480876497886\\
91	-40.447354743375\\
92	-40.4466284788313\\
93	-40.4459087962043\\
94	-40.445195636082\\
95	-40.4444889395876\\
96	-40.4437886483766\\
97	-40.4430947046275\\
98	-40.44240705104\\
99	-40.4417256308289\\
100	-40.4410503877181\\
101	-40.4403812659371\\
102	-40.4397182102167\\
103	-40.439061165782\\
104	-40.4384100783507\\
105	-40.4377648941274\\
106	-40.4371255597986\\
107	-40.4364920225296\\
108	-40.4358642299587\\
109	-40.4352421301948\\
110	-40.4346256718121\\
111	-40.4340148038456\\
112	-40.4334094757877\\
113	-40.432809637584\\
114	-40.432215239629\\
115	-40.431626232763\\
116	-40.4310425682662\\
117	-40.4304641978574\\
118	-40.4298910736878\\
119	-40.4293231483381\\
120	-40.4287603748157\\
121	-40.4282027065487\\
122	-40.4276500973839\\
123	-40.4271025015829\\
124	-40.4265598738181\\
125	-40.4260221691679\\
126	-40.4254893431153\\
127	-40.4249613515436\\
128	-40.4244381507313\\
129	-40.423919697351\\
130	-40.4234059484634\\
131	-40.4228968615164\\
132	-40.4223923943396\\
133	-40.4218925051425\\
134	-40.4213971525088\\
135	-40.420906295397\\
136	-40.420419893132\\
137	-40.4199379054066\\
138	-40.4194602922753\\
139	-40.4189870141519\\
140	-40.4185180318064\\
141	-40.4180533063619\\
142	-40.4175927992915\\
143	-40.4171364724145\\
144	-40.4166842878943\\
145	-40.4162362082344\\
146	-40.4157921962766\\
147	-40.415352215196\\
148	-40.4149162285007\\
149	-40.4144842000265\\
150	-40.4140560939348\\
};
\addlegendentry{\scriptsize \cite{ali2016ofdm} ($\rho = 1$)}

\end{axis}
\end{tikzpicture}%

%% file: figures/SFDR_N_combined_400
%
%
\definecolor{mycolor1}{rgb}{0.00000,0.44700,0.74100}%
\definecolor{mycolor2}{rgb}{0.85000,0.32500,0.09800}%
\definecolor{mycolor3}{rgb}{0.92900,0.69400,0.12500}%
\definecolor{mycolor4}{rgb}{0.49400,0.18400,0.55600}%
\definecolor{mycolor5}{rgb}{0.46600,0.67400,0.18800}%
\definecolor{mycolor6}{rgb}{1.00000,0.00000,1.00000}%
\begin{tikzpicture}
  \begin{groupplot}[
    group style={group size=2 by 1, horizontal sep=1.6cm},
    width=0.45\textwidth,
    height=0.4\textwidth,
    grid=major,
    legend pos=north west
  ]

  \nextgroupplot[
    width=.49\textwidth, 
    height = .4\textwidth, 
    grid, 
    xmin=20,
    xmax=50,
    xlabel={SNR (dB)},
    ymin=-90,
    ymax=-50,
    ylabel={SFDR$_{\text{N}}$ (dB)},
    title={3GPP-based IQ imbalance},
    legend pos=south east, 
    legend cell align=left,
    legend columns=3, 
    legend style={
        at={(0.22,1.3)}, 
        anchor=south west, 
        legend cell align=left, 
        align=left, 
        draw=white!15!black}
    ]
    \addplot [color=red, line width=1.5pt]
      table[row sep=crcr]{%
    20	-57.4969658226006\\
    25	-62.4576363876909\\
    30	-67.5183482750508\\
    35	-72.5548581220162\\
    40	-77.3676311931448\\
    45	-82.4093754825894\\
    50	-87.2240330199847\\
    };
    \addlegendentry{no IQ imb.}

    \addplot [color=black, line width=1.5pt]
      table[row sep=crcr]{%
    20	-56.5153316731052\\
    25	-59.9924035440004\\
    30	-60.4565813416346\\
    35	-64.2483057438294\\
    40	-64.5367861994321\\
    45	-66.318436250057\\
    50	-64.8265345222032\\
    };
    \addlegendentry{with IQ imb.}

    \addplot [color=mycolor1, dashed, line width=1.5pt]
      table[row sep=crcr]{%
    20	-56.2013660623333\\
    25	-61.1738920151247\\
    30	-66.4325207295366\\
    35	-71.6003954142704\\
    40	-76.7369174788929\\
    45	-81.6307135993526\\
    50	-86.0076529725404\\
    };
    \addlegendentry{AWF}

    \addplot [color=mycolor2, dashed, line width=1.5pt]
      table[row sep=crcr]{%
    20	-57.3936739790419\\
    25	-62.4060627394069\\
    30	-67.4041908545023\\
    35	-72.350880659321\\
    40	-77.3284644017588\\
    45	-82.3023357163789\\
    50	-86.9477368794382\\
    };
    \addlegendentry{IWF}

    \addplot [color=mycolor3, dashed, line width=1.5pt]
      table[row sep=crcr]{%
    20	-56.3262799840329\\
    25	-61.5561683945513\\
    30	-66.705325284598\\
    35	-71.3174833533949\\
    40	-75.6187755380242\\
    45	-79.0254898541174\\
    50	-81.0078392685966\\
    };
    \addlegendentry{LMS}

    \addplot [color=mycolor4, dashed, line width=1.5pt]
      table[row sep=crcr]{%
    20	-54.4391547641764\\
    25	-60.4177299961697\\
    30	-65.2202723082513\\
    35	-70.4504346007506\\
    40	-75.3834416748511\\
    45	-80.1550713535568\\
    50	-84.7852146591323\\
    };
    \addlegendentry{NLMS}

    \addplot [color=mycolor5, dashed, line width=1.5pt]
      table[row sep=crcr]{%
    20	-54.620402397598\\
    25	-60.3763702473339\\
    30	-65.6224578069664\\
    35	-70.4052705866443\\
    40	-75.4992156323634\\
    45	-80.1126792971871\\
    50	-84.4148257462198\\
    };
    \addlegendentry{RLS}

    \addplot [color=mycolor6, line width=1.5pt]
      table[row sep=crcr]{%
    20	-55.7316439962765\\
    25	-61.2886814561265\\
    30	-64.7208463189127\\
    35	-68.9148119796792\\
    40	-70.3310769770133\\
    45	-71.1365447642066\\
    50	-72.6073210622797\\
    };
    \addlegendentry{\cite{ali2016ofdm} ($\rho$ = 1)}

    \nextgroupplot[
      width=.49\textwidth, 
      height = .4\textwidth, 
      grid, 
      xmin=20,
      xmax=50,
      xlabel={SNR (dB)},
      ymin=-90,
      ymax=-40,
      title={Literature-based IQ imbalance}
    ]

    \addplot [color=red, line width=1.5pt]
    table[row sep=crcr]{%
    20	-57.5001254061401\\
    25	-62.4979347985118\\
    30	-67.4115915089774\\
    35	-72.4669598777488\\
    40	-77.4472506130131\\
    45	-82.4261382713677\\
    50	-87.1771621943272\\
    };

    \addplot [color=black, line width=1.5pt]
    table[row sep=crcr]{%
    20	-44.5049145716259\\
    25	-41.9380766809899\\
    30	-42.7737444567896\\
    35	-43.8352307115336\\
    40	-45.8214904169102\\
    45	-41.6051038357003\\
    50	-45.5788622686196\\
    };

    \addplot [color=mycolor1, dashed, line width=1.5pt]
    table[row sep=crcr]{%
    20	-52.7028607500108\\
    25	-57.8682144620649\\
    30	-62.3897779588604\\
    35	-68.1039946943211\\
    40	-71.9120706099945\\
    45	-77.3680883035511\\
    50	-76.8142166274512\\
    };

    \addplot [color=mycolor2, dashed, line width=1.5pt]
    table[row sep=crcr]{%
    20	-53.2904587918362\\
    25	-59.1968429716869\\
    30	-63.7485766103515\\
    35	-67.8961890242825\\
    40	-73.1860445586754\\
    45	-77.3930559423518\\
    50	-77.7650820574536\\
    };

    \addplot [color=mycolor3, dashed, line width=1.5pt]
    table[row sep=crcr]{%
    20	-51.9883346451179\\
    25	-56.5267947614878\\
    30	-59.5931832029549\\
    35	-61.6197389649644\\
    40	-64.3436800149767\\
    45	-62.2702004523146\\
    50	-65.3718865969638\\
    };

    \addplot [color=mycolor4, dashed, line width=1.5pt]
    table[row sep=crcr]{%
    20	-50.6910045201419\\
    25	-54.5062644544407\\
    30	-59.0260598094852\\
    35	-64.1830665690233\\
    40	-68.205040767841\\
    45	-62.718835584815\\
    50	-72.6317182883585\\
    };

    \addplot [color=mycolor5, dashed, line width=1.5pt]
    table[row sep=crcr]{%
    20	-51.6221911322755\\
    25	-56.5557630618148\\
    30	-60.3686153861183\\
    35	-65.5739537442306\\
    40	-69.1219925447917\\
    45	-67.152094046061\\
    50	-73.1765614790853\\
    };

    \addplot [color=mycolor6, line width=1.5pt]
    table[row sep=crcr]{%
    20	-54.5981633820879\\
    25	-59.1417478895138\\
    30	-63.6462871120121\\
    35	-65.8736919931479\\
    40	-67.7284674886387\\
    45	-69.3950121832741\\
    50	-68.3342753235986\\
    };
  \end{groupplot}
\end{tikzpicture}%

%% file: figures/SFDR_G_combined_400
%
%
\definecolor{mycolor1}{rgb}{0.00000,0.44700,0.74100}%
\definecolor{mycolor2}{rgb}{0.85000,0.32500,0.09800}%
\definecolor{mycolor3}{rgb}{0.92900,0.69400,0.12500}%
\definecolor{mycolor4}{rgb}{0.49400,0.18400,0.55600}%
\definecolor{mycolor5}{rgb}{0.46600,0.67400,0.18800}%
\definecolor{mycolor6}{rgb}{1.00000,0.00000,1.00000}%
\begin{tikzpicture}
  \begin{groupplot}[
    group style={group size=2 by 1, horizontal sep=1.6cm},
    width=0.45\textwidth,
    height=0.4\textwidth,
    grid=major,
    legend pos=north west
  ]

  \nextgroupplot[
    width=.49\textwidth, 
    height = .4\textwidth, 
    grid, 
    xmin=20,
    xmax=50,
    xlabel={SNR (dB)},
    ymin=-100,
    ymax=-55,
    ylabel={SFDR$_{\text{G}}$ (dB)},
    title={3GPP-based IQ imbalance},
    legend pos=south east, 
    legend cell align=left,
    legend columns=3, 
    legend style={
        at={(0.22,1.3)}, 
        anchor=south west, 
        legend cell align=left, 
        align=left, 
        draw=white!15!black}
    ]
    \addplot [color=red, line width=1.5pt]
      table[row sep=crcr]{%
    20	-66.620556547399\\
    25	-71.6577780816168\\
    30	-76.4148512010088\\
    35	-81.5567168579619\\
    40	-86.3292602253962\\
    45	-91.2017922193679\\
    50	-95.3090707132562\\
    };
    \addlegendentry{no IQ imb.}

    \addplot [color=black, line width=1.5pt]
      table[row sep=crcr]{%
    20	-65.4054484477149\\
    25	-69.2331242660017\\
    30	-69.4843269245934\\
    35	-73.1638365181043\\
    40	-73.75033209352\\
    45	-75.4163077554266\\
    50	-73.7805865554829\\
    };
    \addlegendentry{with IQ imb.}

    \addplot [color=mycolor1, dashed, line width=1.5pt]
      table[row sep=crcr]{%
    20	-65.3171785547673\\
    25	-70.5157719437374\\
    30	-75.0560113043091\\
    35	-80.781966605715\\
    40	-85.2911731183558\\
    45	-89.5889257240728\\
    50	-92.6594731820049\\
    };
    \addlegendentry{AWF}

    \addplot [color=mycolor2, dashed, line width=1.5pt]
      table[row sep=crcr]{%
    20	-66.2367738833457\\
    25	-71.6688453815476\\
    30	-76.0507399337243\\
    35	-81.3126374936134\\
    40	-86.2910576285377\\
    45	-91.0063974202642\\
    50	-95.2062919339572\\
    };
    \addlegendentry{IWF}

    \addplot [color=mycolor3, dashed, line width=1.5pt]
      table[row sep=crcr]{%
    20	-64.3857424652941\\
    25	-69.0835467974529\\
    30	-74.5661864291065\\
    35	-79.4848892244567\\
    40	-83.9198386824344\\
    45	-87.114485706966\\
    50	-89.1015485516669\\
    };
    \addlegendentry{LMS}

    \addplot [color=mycolor4, dashed, line width=1.5pt]
      table[row sep=crcr]{%
    20	-61.7062872085727\\
    25	-68.0357155533216\\
    30	-72.2831156172544\\
    35	-77.7669174518816\\
    40	-82.5705691803514\\
    45	-87.7444337691925\\
    50	-91.6598294409402\\
    };
    \addlegendentry{NLMS}

    \addplot [color=mycolor5, dashed, line width=1.5pt]
      table[row sep=crcr]{%
    20	-60.3594927090928\\
    25	-63.3453788278702\\
    30	-71.6979710233203\\
    35	-73.4424938985153\\
    40	-78.2688849808686\\
    45	-73.2527534801683\\
    50	-76.7544584962081\\
    };
    \addlegendentry{RLS}

    \addplot [color=mycolor6, line width=1.5pt]
      table[row sep=crcr]{%
    20	-68.5832923117428\\
    25	-68.3994265196931\\
    30	-74.7030665928805\\
    35	-79.1073885776741\\
    40	-81.7197903942516\\
    45	-80.5329589277974\\
    50	-81.9982453560282\\
    };
    \addlegendentry{\cite{ali2016ofdm} ($\rho$ = 1)}

    \nextgroupplot[
      width=.49\textwidth, 
      height = .4\textwidth, 
      grid, 
      xmin=20,
      xmax=50,
      xlabel={SNR (dB)},
      ymin=-100,
      ymax=-45,
      title={Literature-based IQ imbalance}
    ]
    \addplot [color=red, line width=1.5pt]
    table[row sep=crcr]{%
    20	-66.4233087284028\\
    25	-71.2652540292108\\
    30	-76.481645448653\\
    35	-81.180935638373\\
    40	-86.4401424210942\\
    45	-91.3926669300527\\
    50	-95.9159605128211\\
    };

    \addplot [color=black, line width=1.5pt]
    table[row sep=crcr]{%
    20	-52.5574913609437\\
    25	-50.4439532136654\\
    30	-51.3834860223335\\
    35	-52.3155504350947\\
    40	-53.6782578315384\\
    45	-50.3274501792018\\
    50	-54.1670884631205\\
    };

    \addplot [color=mycolor1, dashed, line width=1.5pt]
    table[row sep=crcr]{%
    20	-56.9803171567514\\
    25	-57.1537118097552\\
    30	-59.9897343354345\\
    35	-61.4322767648564\\
    40	-58.5541119192922\\
    45	-64.124431866627\\
    50	-58.6616557896253\\
    };

    \addplot [color=mycolor2, dashed, line width=1.5pt]
    table[row sep=crcr]{%
    20	-59.5320756195071\\
    25	-62.6123072916857\\
    30	-61.3848574898904\\
    35	-61.6605695827859\\
    40	-62.2921990866971\\
    45	-64.4689049056345\\
    50	-59.8911933448629\\
    };

    \addplot [color=mycolor3, dashed, line width=1.5pt]
    table[row sep=crcr]{%
    20	-57.3374107077516\\
    25	-60.7060948874185\\
    30	-60.3041190545698\\
    35	-60.8004282289454\\
    40	-61.0505097205896\\
    45	-62.9847843585493\\
    50	-58.9800549998471\\
    };

    \addplot [color=mycolor4, dashed, line width=1.5pt]
    table[row sep=crcr]{%
    20	-54.8572050939169\\
    25	-56.2838332320034\\
    30	-56.5060907695315\\
    35	-56.8963749003571\\
    40	-57.1303418860607\\
    45	-57.0002592082856\\
    50	-56.9354076305839\\
    };

    \addplot [color=mycolor5, dashed, line width=1.5pt]
    table[row sep=crcr]{%
    20	-52.1653162444247\\
    25	-52.490965044927\\
    30	-53.9064117232457\\
    35	-55.3739078310662\\
    40	-56.0162049581132\\
    45	-53.9245570530629\\
    50	-55.9606517912751\\
    };

    \addplot [color=mycolor6, line width=1.5pt]
    table[row sep=crcr]{%
    20	-59.0250060194188\\
    25	-57.7742416125152\\
    30	-58.3139690761175\\
    35	-59.8193673051295\\
    40	-60.3999768243533\\
    45	-58.288518529224\\
    50	-60.6273169851739\\
    };
  \end{groupplot}
\end{tikzpicture}%